\documentclass[runningheads]{llncs}
\usepackage{graphicx}

\usepackage{float}
\usepackage{tikz}
\usepackage{comment}
\usepackage{amsmath,amssymb} 
\usepackage{color, colortbl}
\usepackage{adjustbox}
\usepackage{array, enumitem, soul, algorithm, algpseudocode}
\usepackage{graphicx}
\usepackage{subcaption}
\usepackage{cite}
\usepackage{xcolor}
\usepackage{xspace}
\usepackage{booktabs}
\usepackage{pifont}
\usepackage[pagebackref,breaklinks,colorlinks]{hyperref}

\usepackage[accsupp]{axessibility}  

\newcommand{\cmark}{\text{\ding{51}}}%
\newcommand{\xmark}{\text{\ding{55}}}%

\definecolor{Gray}{gray}{0.92}

\makeatletter

\DeclareRobustCommand\onedot{\futurelet\@let@token\@onedot}
\def\@onedot{\ifx\@let@token.\else.\null\fi\xspace}

\def\eg{\emph{e.g}\onedot} 
\def\ie{\emph{i.e}\onedot}

\def\et{\emph{et al}\onedot}
\newcommand{\printfnsymbol}[1]{%
  \textsuperscript{1}
}
\makeatother

\begin{document}
\pagestyle{headings}
\mainmatter
\def\ECCVSubNumber{100}  

\title{Reversed Image Signal Processing and RAW Reconstruction. AIM 2022 Challenge Report} 

\titlerunning{AIM 2022 Reversed ISP Challenge}
%
\author{Marcos~V.~Conde\inst{1} \and Radu Timofte\inst{1} \and
Yibin Huang \and Jingyang Peng \and Chang Chen \and Cheng Li \and Eduardo P{\'e}rez-Pellitero \and Fenglong Song \and
Furui Bai \and Shuai Liu \and Chaoyu Feng \and Xiaotao Wang \and Lei Lei \and
Yu Zhu \and Chenghua Li \and Yingying Jiang \and Yong A \and Peisong Wang \and Cong Leng \and Jian Cheng \and
Xiaoyu Liu \and Zhicun Yin \and Zhilu Zhang \and Junyi Li \and Ming Liu \and Wangmeng Zuo \and
Jun Jiang \and Jinha Kim \and
Yue Zhang \and Beiji Zou \and
Zhikai Zong \and Xiaoxiao Liu \and
Juan Mar{\'i}n Vega \and Michael Sloth \and Peter Schneider-Kamp \and Richard Röttger \and
Furkan Kınlı \and Barış Özcan \and Furkan Kıraç \and
Li Leyi \and
SM Nadim Uddin \and Dipon Kumar Ghosh \and Yong Ju Jung
}
\authorrunning{Conde and Timofte et al.}

\institute{\printfnsymbol{1}~Organizers. Computer Vision Lab, CAIDAS, University of Würzburg, Germany
\email{\{marcos.conde-osorio,radu.timofte\}@uni-wuerzburg.de}\\
\url{https://data.vision.ee.ethz.ch/cvl/aim22/}\\
\url{https://github.com/mv-lab/AISP/}
}

\maketitle


\begin{abstract}
Cameras capture sensor RAW images and transform them into pleasant RGB images, suitable for the human eyes, using their integrated Image Signal Processor (ISP).
Numerous low-level vision tasks operate in the RAW domain (\eg{}~image denoising, white balance) due to its linear relationship with the scene irradiance, wide-range of information at 12bits, and sensor designs. Despite this, RAW image datasets are scarce and more expensive to collect than the already large and public RGB datasets.
This paper introduces the AIM 2022 Challenge on Reversed Image Signal Processing and RAW Reconstruction. We aim to recover raw sensor images from the corresponding RGBs without metadata and, by doing this, ``reverse" the ISP transformation. The proposed methods and benchmark establish the state-of-the-art for this low-level vision inverse problem, and generating realistic raw sensor readings can potentially benefit other tasks such as denoising and super-resolution.
\end{abstract}

\section{Introduction}

The majority of low-level vision and computational photography tasks use RGB images obtained from the in-camera Image Signal Processor (ISP)~\cite{conde2022model} that converts the camera sensor's raw readings into perceptually pleasant RGB images, suitable for the human visual system. One of the reasons is the accessibility and amount of RGB datasets. Multiple approaches have been proposed to model the RAW to RGB transformation (\ie ISP) using deep neural networks. We can highlight FlexISP~\cite{heide2014flexisp}, DeepISP~\cite{schwartz2018deepisp} and PyNET~\cite{Ignatov_2020_pynet, ignatov2020aim, ignatov2022isp}.

However, the characteristics of raw camera's sensor data (\eg linear relationship with scene irradiance at each in 12-14 bits, unprocessed signal and noise samples) are often better suited for the inverse problems that commonly arise in low-level vision tasks such as denoising, deblurring, super-resolution~\cite{abdelhamed2019ntire, bhat2021ntire, qian2019trinity, gharbi2016deep}. Professional photographers also commonly choose to process RAW images by themselves to produce images with better visual effects~\cite{karaimer2016software}.
Unfortunately, RAW image datasets are not as common and diverse as their RGB counterparts, therefore, CNN-based approaches might not reach their full potential. To bridge this gap, we introduce the AIM Reversed ISP Challenge and review current solutions.
We can find metadata-based raw reconstruction methods for de-render or unprocess the RGB image back to its original raw values~\cite{Nguyen_2016_CVPR, punnappurath2021spatially,Nam_2022_CVPR}, these methods usually require specific metadata stored as an overhead (\eg{} 64 KB).
Similarly, UPI~\cite{brooks2019unprocessing} proposes a generic camera ISP model composed of five canonical and invertible steps~\cite{delbracio2021mobile, karaimer2016software}. This simple approach can produce realistic raw sensor data for training denoising models~\cite{brooks2019unprocessing}. However, it requires specific camera parameters (\eg correction matrices, digital gains) that are generally inaccessible.

Learning-based approaches~\cite{punnappurath2019learning, zamir2020cycleisp,xing2021invertible, conde2022model, afifi2021cie} attempt to learn the ISP, and how to reverse it, in an end-to-end manner.
CycleISP~\cite{zamir2020cycleisp} uses a 2-branch model to learn the RAW2RGB and RGB2RAW transformations. InvISP~\cite{xing2021invertible} models the camera ISP as an invertible function using normalizing-flows~\cite{kobyzev2020normalizing}. These learned methods are essentially non-interpretable black-boxes. MBISPLD~\cite{conde2022model} proposes a novel learnable, invertible and interpretable model-based ISP, based on classical designs~\cite{delbracio2021mobile, karaimer2016software}.
In a more complex domain, Arad \et{} proposed the the recovery of whole-scene hyperspectral (HS) information from a RGB image~\cite{arad2020ntire}.


Despite other approaches explored how to recover RAW information~\cite{punnappurath2019learning, Nguyen_2016_CVPR, punnappurath2021spatially, brooks2019unprocessing}, to the best of our knowledge, this is the first work to analyze exhaustively this novel (or at least, not so studied) inverse problem. 

\textbf{Our contribution.} In this paper, we introduce the first benchmark for learned raw sensor image reconstruction. We use data from 4 different smartphone sensors and analyze 14 different learned methods (\ie no metadata or prior information about the camera is required).

This challenge is one of the AIM 2022 associated challenges: reversed ISP~\cite{conde2022aim}, efficient learned ISP~\cite{ignatov2022isp}, super-resolution of compressed image and video~\cite{yang2022aim}, efficient image super-resolution~\cite{ignatov2022isr}, efficient video super-resolution~\cite{ignatov2022vsr}, efficient Bokeh effect rendering~\cite{ignatov2022bokeh}, efficient monocular depth estimation~\cite{ignatov2022depth}, Instagram filter removal~\cite{kinli2022aim}.

\vspace{-4mm}

\begin{figure}[!hb]
\centering
\includegraphics[width=\linewidth]{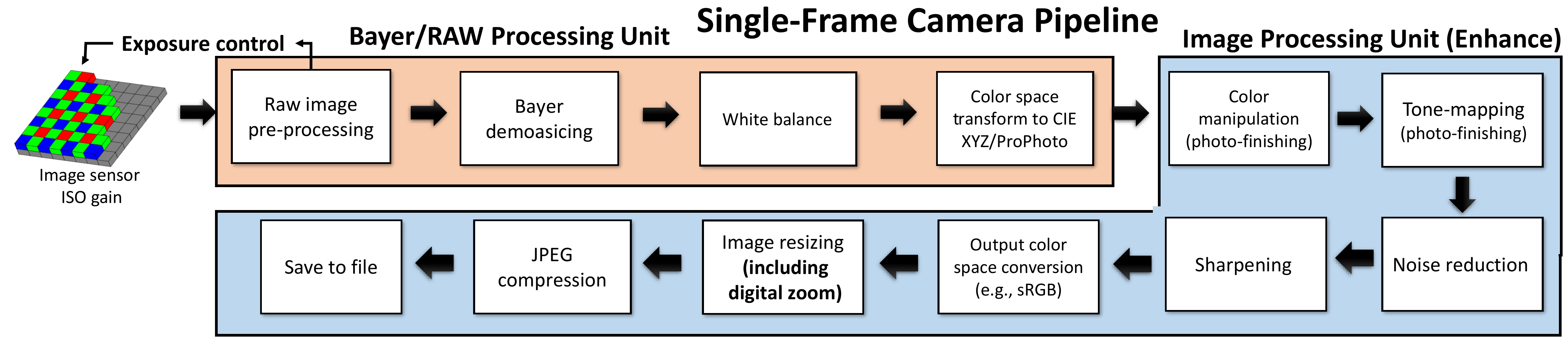}
\caption{Classical Image Signal Processor (ISP). Image from Delbracio \et{}~\cite{delbracio2021mobile}.}
\label{fig:teaser}
\end{figure}

\clearpage

\section{AIM 2022 Reversed ISP Challenge}

The objectives of the AIM 2022 Reversed ISP Challenge are: (i) to propose the first challenge and benchmark for RAW image reconstruction and Reversed Image Signal Processing; (ii) to establish the state-of-the-art in this low-level vision inverse problem; (iii) to compare different contemporary solutions and analyze their drawbacks.
The challenge consists on two tracks: Track 1 for the analysis of the \emph{Samsung S7} sensor (Sony IMX260); Track 2 for the analysis of the \emph{Huawei P20 Pro} sensor (Sony IMX380). Both tracks are structured in the same way: participants use the provided training datasets and baseline~\footnote{\url{https://github.com/mv-lab/AISP/}}, and submit their results (\ie generated RAW images) through an online server with a public leaderboard~\footnote{\url{https://codalab.lisn.upsaclay.fr/competitions/5079}} where they can compare with other methods. Finalists must submit reproducible code and a detailed factsheet describing their solution.
\textbf{Overview:} 80 unique participants registered at the challenge, with 11 teams competing in the final testing phase. More than 200 submissions were received.

\subsection{Datasets}
We aim to promote research and reproducibility in this area, for this reason, we select public available datasets for training and testing the methods. Moreover, we select smartphone's datasets, as these are, by design, more complex ISPs than DSLR cameras~\cite{delbracio2021mobile, karaimer2016software} and therefore more challenging to model and ``reverse".

\begin{itemize}
    \item Track 1. Samsung \textbf{S7} Dataset from DeepISP~\cite{schwartz2018deepisp}. Raw and processed RGB images of the 110 different scenes captured by the phone camera with normal exposure time. The images have resolution $3024\times4032$ (12MP Sony IMX260). Image pairs are aligned enough to use pixel-wise metrics.
    \vspace{2mm}
    \item Track 2. ETH Huawei \textbf{P20} Pro Dataset (PyNET)~\cite{Ignatov_2020_pynet}. RAW and RGB images obtained with Huawei's built-in ISP (12.3 MP Sony Exmor IMX380). The photos were captured in automatic mode, and default settings were used throughout the whole collection procedure. Original images have resolution $3840\times5120$. However, the pairs of raw sensor data and RGB have a notable misalignment (over 8px).
    \vspace{2mm}
    \item SSID Dataset~\cite{abdelhamed2018sidd} clean RAW-RGB image pairs of the Samsung Galaxy \textbf{S6} Edge phone (16MP Sony IMX240), captured under normal brightness.
    \vspace{2mm}
    \item RAW-to-RAW Dataset~\cite{afifi2021semi}. Authors provide natural scene RAW and camera-ISP rendered RGB pairs using the Samsung Galaxy \textbf{S9} (Sony IMX345).
\end{itemize}

Datasets are manually filtered. Considering the original full-resolution (FR) images, we unify RAWs to a common RGGB pattern, next, we extract non-overlapping crops ($504\times504$ in track 1, $496\times496$ in track 2) and save them as 4-channel 10-bit images in \texttt{.npy} format. Some teams used directly FR images.

\subsection{Evaluation and Results}
\label{sec:eval}

We use 20\% of the S6 dataset~\cite{schwartz2018deepisp} and P20~\cite{Ignatov_2020_pynet} as validation and test sets for each track. Since the provided S6 and P20 training and testing (``Test1") sets are public, we use an additional internal test set (``Test2") not disclosed until the end of the challenge. For the Track 1 (S7), the internal test contains additional images from the previously explained S9~\cite{afifi2021semi} and S6~\cite{abdelhamed2018sidd} datasets. For the Track 2 (P20) we use extra noisy images from~\cite{Ignatov_2020_pynet}. This allows us to evaluate the generalization capabilities and robustness of the proposed solutions.

Table~\ref{tab:bench} represents the challenge benchmark~\cite{conde2022aim}. The best 3 approaches achieve above 30dB PSNR, which denotes a great reconstruction of the original RAW readings captured by the camera sensor. Attending to ``Test2", the proposed methods can generalize and produce realistic RAW data for unseen similar sensors. For instance, models trained using S7~\cite{schwartz2018deepisp} data, can produce realistic RAWs for S9~\cite{afifi2021semi} and S6~\cite{abdelhamed2018sidd} (previous and next-generation sensors).

Attending to the PSNR and SSIM values, we can see a clear fidelity-perception tradeoff~\cite{blau2018perception} \ie{ at the ``Test2" of Track 2, the top solutions scored the same SSIM and notably different PSNR values}. To explore this, we provide extensive qualitative comparisons of the proposed methods in the Appendix~\ref{app:extra-quali-results}.

\vspace{-5mm}

\begin{table}[!ht]
    \centering
    \caption{AIM Reversed ISP Challenge Benchmark. Teams are ranked based on their performance on \underline{Test1} and \underline{Test2}, an internal test set to evaluate the generalization capabilities and robustness of the proposed solutions.
    The methods (*) have trained using extra data from~\cite{schwartz2018deepisp}, and therefore only results on the internal datasets are relevant. CycleISP~\cite{zamir2020cycleisp} was reported by multiple participants
    }
    \label{tab:bench}
    \vspace{2mm}
    \resizebox{\linewidth}{!}{
    \begin{tabular}{l||c|c||c|c||c|c||c|c}
        \hline\noalign{\smallskip}
        & \multicolumn{4}{c ||}{\textbf{Track 1 (Samsung S7)}} & \multicolumn{4}{c}{\textbf{Track 2 (Huawei P20)}} \\
        Team & \multicolumn{2}{c||}{Test1} & \multicolumn{2}{c||}{Test2} & \multicolumn{2}{c||}{Test1} & \multicolumn{2}{c}{Test2} \\
         name & PSNR~$\uparrow$ & SSIM~$\uparrow$ & PSNR~$\uparrow$ & SSIM~$\uparrow$ & PSNR~$\uparrow$ & SSIM~$\uparrow$ & PSNR~$\uparrow$ & SSIM~$\uparrow$ \\
        \hline
        \rowcolor{Gray} NOAHTCV	 & 31.86 & 0.83 & 32.69 & 0.88 & 38.38 & 0.93 & 35.77 & 0.92 \\
        MiAlgo	    & 31.39 & 0.82 & 30.73 & 0.80 & 40.06 & 0.93 & 37.09 & 0.92  \\
        \rowcolor{Gray} CASIA LCVG (*) & 30.19 & 0.81 & 31.47 & 0.86 & 37.58 & 0.93 & 33.99 & 0.92  \\
        HIT-IIL	    & 29.12 & 0.80 & 30.22 & 0.87  & 36.53 & 0.91 & 34.25 & 0.90 \\
        \rowcolor{Gray} SenseBrains	& 28.36 & 0.80 & 30.08 & 0.86 & 35.47 & 0.92 & 32.63 & 0.91 \\
        CS\textasciicircum2U (*)  & 29.13 & 0.79 & 29.95 & 0.84 & - & - & - & -\\
        \rowcolor{Gray}HiImage	    & 27.96 & 0.79 & - & - & 34.40 & 0.94 & 32.13 & 0.90 \\
        0noise	    & 27.67 & 0.79 & 29.81 & 0.87 & 33.68 & 0.90 & 31.83 & 0.89 \\
        \rowcolor{Gray} OzU VGL	    & 27.89 & 0.79 & 28.83 & 0.83 & 32.72 & 0.87 & 30.69 &0.86 \\
        PixelJump   & 28.15 & 0.80 & - & - & -  & - & - & - \\
        \rowcolor{Gray} CVIP & 27.85 & 0.80 & 29.50 & 0.86 & -  & - & - & - \\
        \hline
        CycleISP~\cite{zamir2020cycleisp}  & 26.75 & 0.78 & - & - & 32.70 & 0.85 & - & -  \\
        \rowcolor{Gray} UPI~\cite{brooks2019unprocessing}  & 26.90 & 0.78 & - & - & - & - & - & -  \\
        U-Net Base & 26.30 & 0.77 & - & - & 30.01 & 0.80 & - & -  \\
        \hline
    \end{tabular}}
\end{table}

In Section~\ref{sec:solutions}, we introduce the solutions from each team. All the proposed methods are learned-based and do not require metadata or specific camera parameters. Multiple approaches can serve as a plug-in augmentation.

\clearpage
\section{Proposed Methods and Teams}
\label{sec:solutions}

The complete information about the teams can be consulted in the Appendix~\ref{app:affiliations}.

\subsection{NOAHTCV}
\emph{Yibin Huang, Jingyang Peng, Chang Chen, Cheng Li, Eduardo P{\'e}rez-Pellitero, and Fenglong Song}
\vspace{2mm}

\begin{figure}[]
\centering
\includegraphics[width=\textwidth, keepaspectratio]{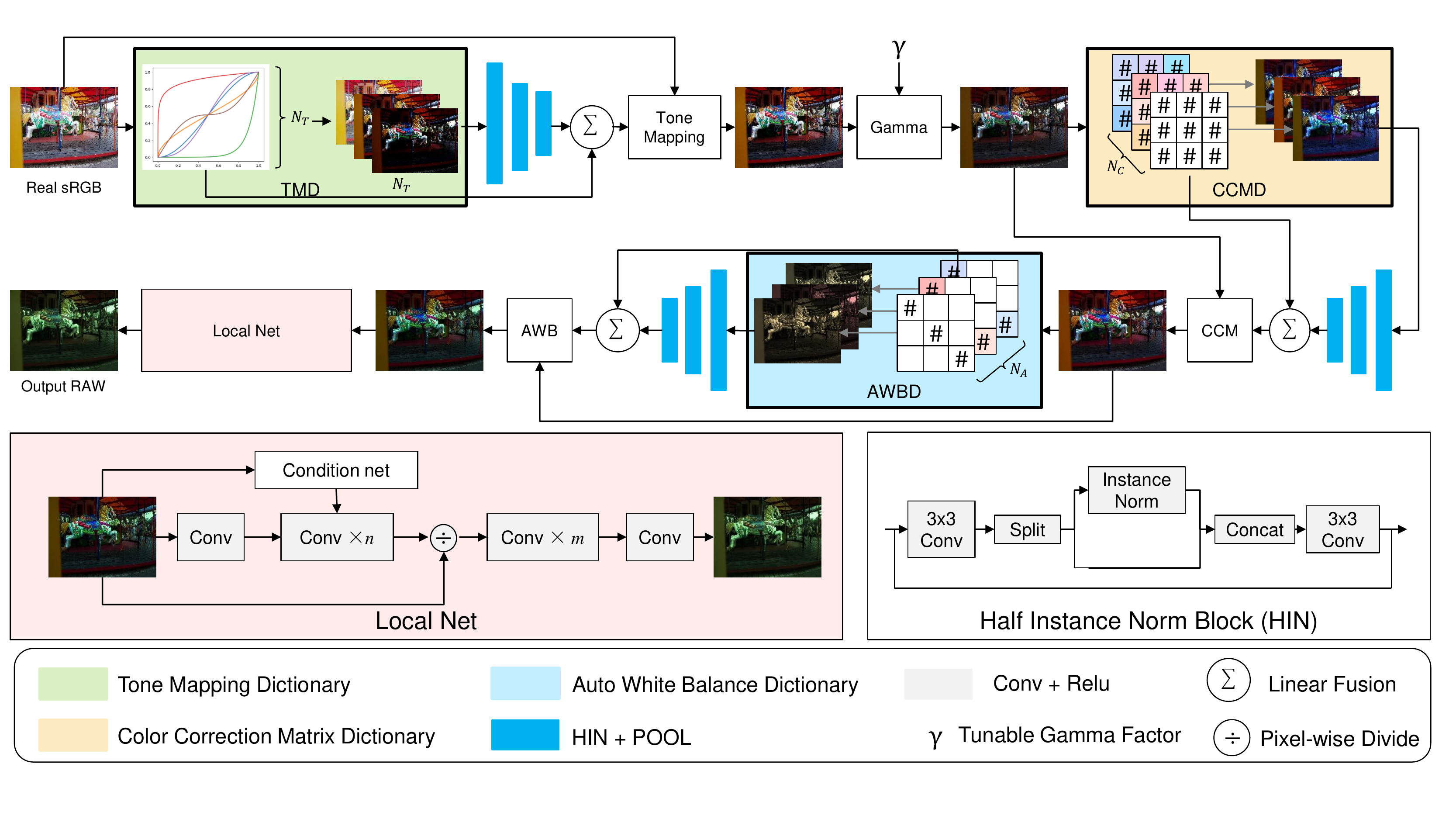} 
\caption{NOAHTCV method overview, inspired in MBISPLD~\cite{conde2022model}.}
\label{fig:team1}
\end{figure}

\noindent This method is illustrated in Figure~\ref{fig:team1}, it is based on MBISPLD~\cite{conde2022model}, a model-based interpretable approach. The only remarkable differences are (i) a modified inverse tone mapping block and (ii) LocalNet, a custom CNN added on top to improve performance in exchange of losing invertibility and efficiency. 
The ``inverse tone mapping" is formulated as as a convex combination of several monotonic curves, parameterized in a polynomial form. Note that these curves are fixed and not learned together with the model.
The ``LocalNet" is designed to predict conditions for feature modulation, and can be viewed as a variant of local mapping. The output maintains fine details and inverts only tone and color. Authors emphasize in their ablation studies that this LocalNet contributes significantly to improve performance. The team adopts Half Instance Normalization (HIN)~\cite{chen2021hinet} as a basic component in this network.
\textbf{Ablation studies:} The original MBISPLD~\cite{conde2022model} achieved 34.02dB using 0.54M parameters (8 GMACs) and a runtime of 14ms, after adding the custom inverse tone mapping and LocalNet, the model can achieve 36.20dB (improvement of 2dB) in exchange of being $\times10$ more complex (80 GMACs, 5.7M parameters) and $\times2$ slower.
Authors also compare with other methods such as UPI~\cite{brooks2019unprocessing}, DRN~\cite{zhang2018residual} and ESRGAN~\cite{wang2018esrgan}; the base method MBISPLD~\cite{conde2022model} and their proposed modification achieved the best results.
\vspace{-4mm}

\clearpage

\subsection{MiAlgo}
\emph{Furui Bai, Shuai Liu, Chaoyu Feng, Xiaotao Wang, Lei Lei}
\vspace{2mm}

\begin{figure}[!ht]
    \centering
    \setlength{\tabcolsep}{2.0pt}
    \begin{tabular}{c}
    \includegraphics[width=0.91\linewidth]{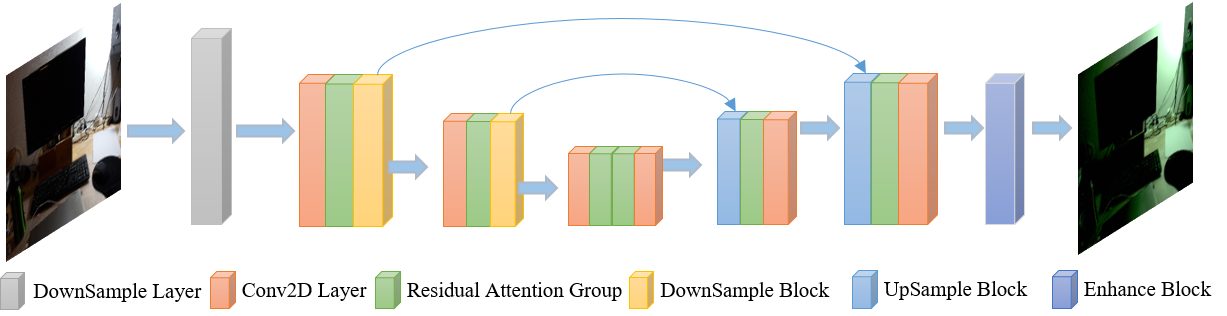} \tabularnewline
    \includegraphics[width=0.9\linewidth]{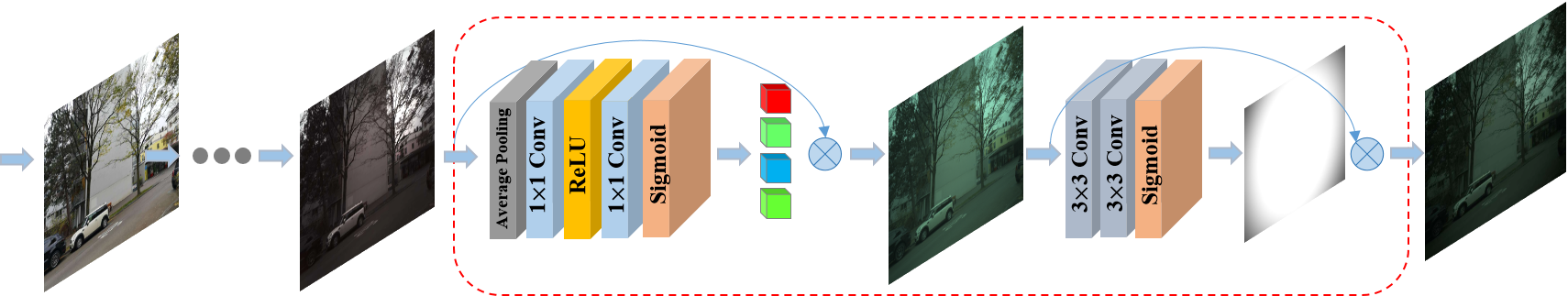}\tabularnewline
    \includegraphics[width=0.9\linewidth]{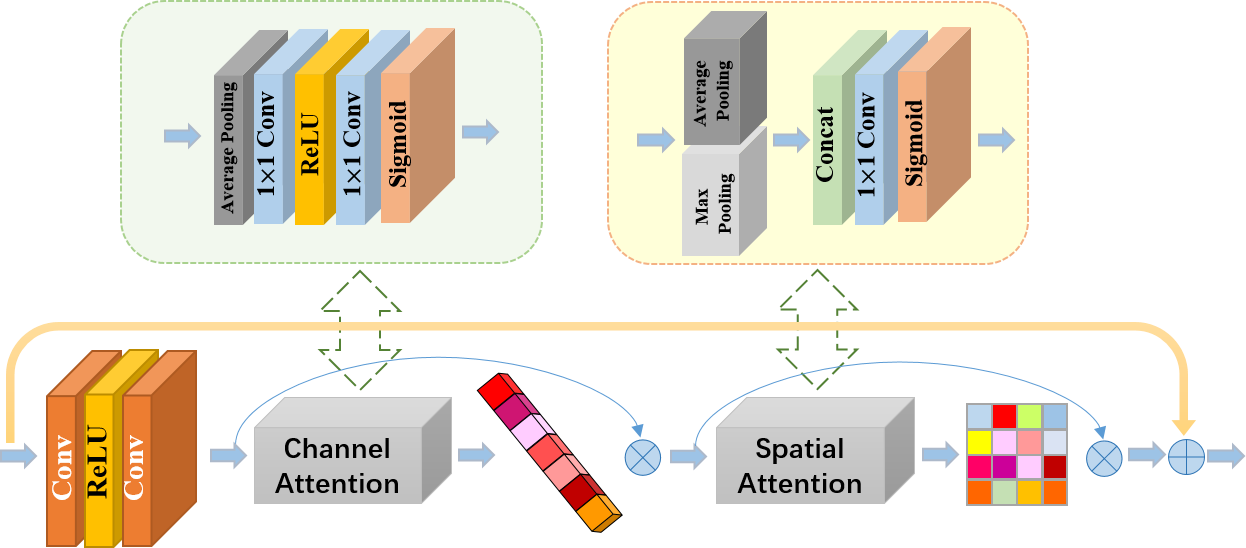}\tabularnewline
    \end{tabular}
    \caption{``Fast and Accurate Network for Reversed ISP" by MiAlgo team. From top to bottom: general framework, enhanced block, residual group.}
    \label{fig:team2}
\end{figure}

\noindent The proposed method has an end-to-end encoder-decoder structure, illustrated in Figure~\ref{fig:team2}.
The approach is a lightweight structure that exploits the powerful capabilities of information encoding-decoding, and therefore, it is able to handle full-resolution (FR) inputs efficiently.
The UNet-like structure consists of many sampling blocks and residual groups to obtain deep features. The main components are the ``residual group"~\cite{zamir2020cycleisp} and the ``enhanced block"~\cite{conde2022model}, these can be visualized Figure~\ref{fig:team2}.
Authors emphasize that since many modules in the ISP are based on full-resolution (\eg{} global tone mapping, auto white balance, and lens shading correction), FR training and testing improves performance in comparison to patch-based approaches.
The method was implemented in PyTorch 1.8, and was trained roughly 13 hrs on 4$\times$ V100 GPU using FR images, L1 loss and SSIM loss, and batch size 1. 
Authors correct misalignments and black leve of RAW images.
We must highlight that the method is able to \textbf{process 4K images in 18ms} in a single V100 GPU. Moreover, it generalizes successfully to noisy inputs and other unseen sensors (see Table~\ref{tab:bench}). The code will be open-sourced.

\vspace{-5mm}

\clearpage
\subsection{CASIA LCVG}
\emph{Yu Zhu,  Chenghua Li, Cong Leng, Jian Cheng}
\vspace{2mm}

\begin{figure}[!b]
\centering
\includegraphics[width=\textwidth]{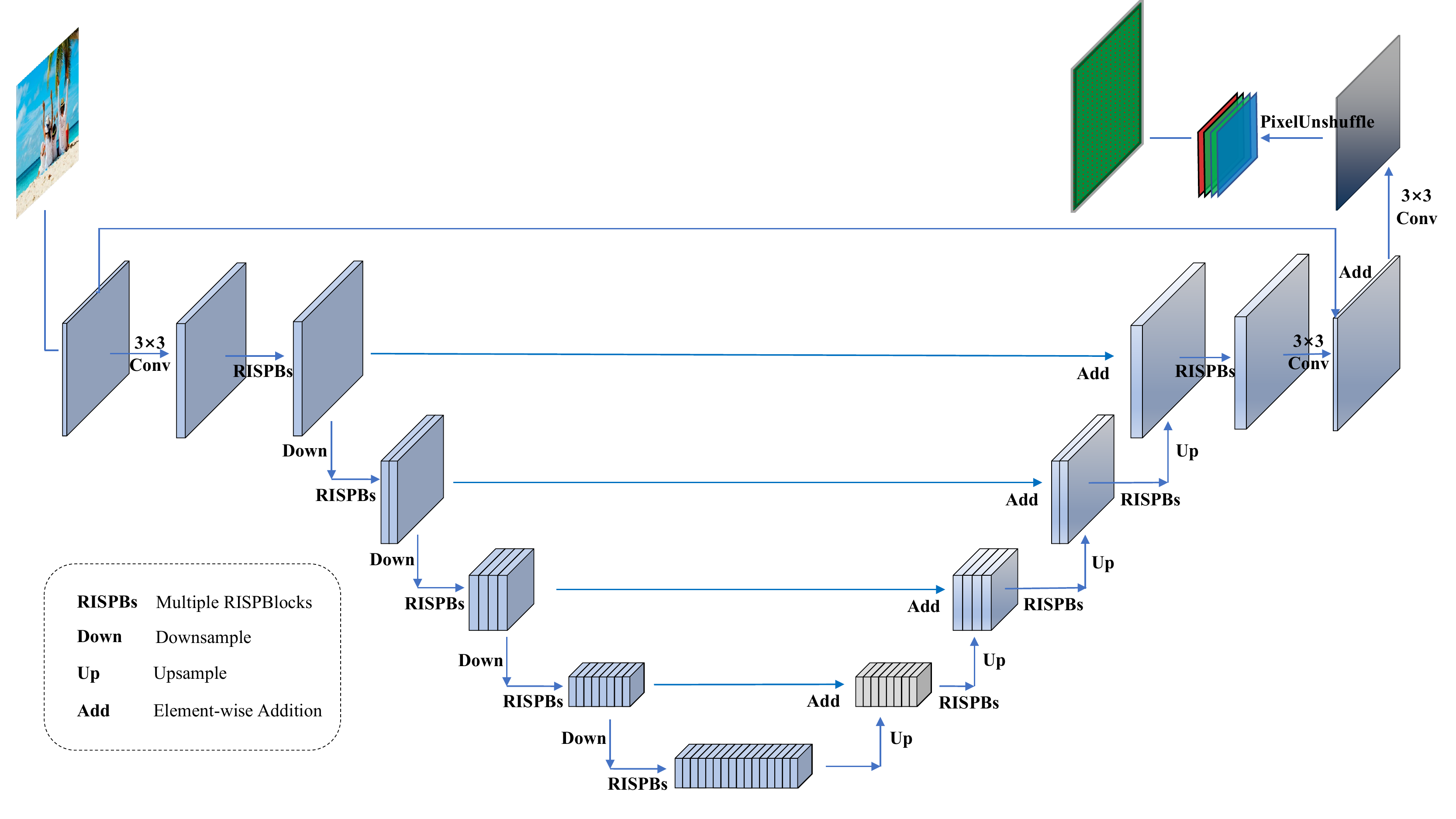}
\caption{RISPNet by CASIA LCVG.}
\label{fig:team3}
\end{figure}

\noindent The team proposes a new reversed ISP network (RISPNet)~\cite{dong2022RISPNet} to achieve efficient RGB to RAW mapping. This is a novel encoder-decoder network with a third-order attention (TOA) module. Since attention~\cite{vaswani2017attention, woo2018cbam} facilitates the complex trade-off between achieving spatial detail and high level contextual information, it can help to recover high dynamic range RAW and thus making the whole recovery process a more manageable step. The architecture of RISPNet is illustrated in Figure~\ref{fig:team3}, it consists of multi-scale hierarchical design incorporating the basic RISPblock, which is a residual attention block combining layer normalization, depth-wise convolutions and channel-spatial attention~\cite{woo2018cbam, chen2022simplenadnet}.

We must note that the team uses the DeepISP S7 dataset~\cite{schwartz2018deepisp}, besides the proposed challenge dataset, this implies that \textbf{the results are not accurate} for this method, since they have potentially trained with extra validation data, nevertheless this should not affect the results in the ``Test2" sets or ``Track2".

The method was implemented in Pytorch 1.10, and was trained roughly 2 days on 8$\times$ A100 GPU using cropped images, AdamW, L1 loss and batch size 32. For data augmentation, the team uses horizontal and vertical flips.

The model is relatively efficient, it can process a 504px RGBs in 219ms using self-ensembles (\ie fuse 8 inputs obtained by flip and rotate operations), which improves their performance by $0.2$dB. Note that this model has 464M parameters, and therefore, it is only suitable for offline RAW synthesis.

\clearpage
\subsection{HIT-IIL}
\emph{Xiaoyu Liu, Zhicun Yin, Zhilu Zhang, Junyi Li, Ming Liu, Wangmeng Zuo}

\vspace{2mm}

\begin{figure}[!b]
\centering
\includegraphics[width=\textwidth]{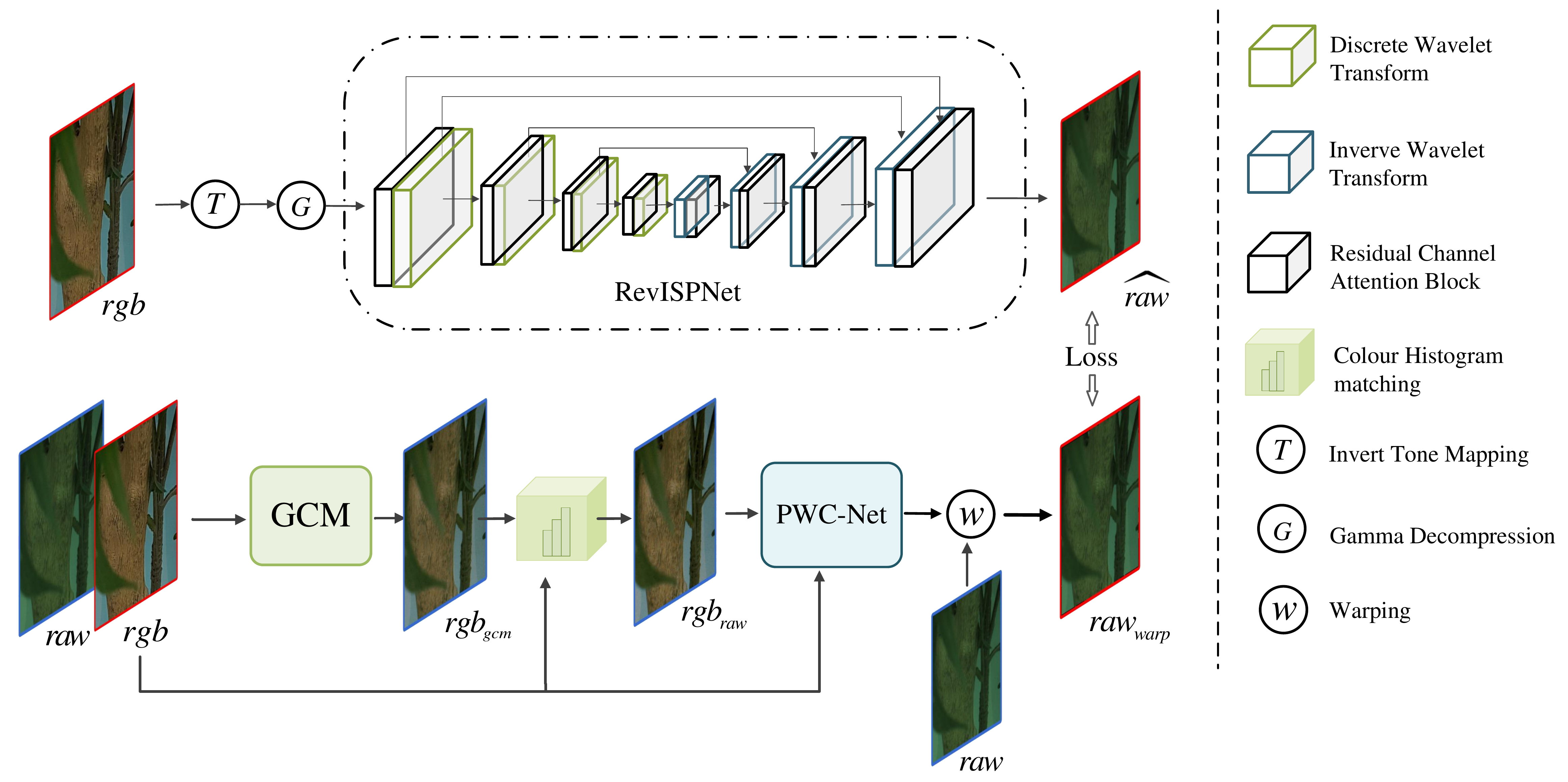}
\caption{Solution from HIT-IIL team, based on LiteISPNet~\cite{zhang2021learningisp} and NAFNet~\cite{chen2022simplenadnet}.}

\label{fig:team4}
\end{figure}

\noindent The team's solution is ``Learning Reverse ISP with Inaccurately Aligned Supervision", inspired by their previous work~\cite{zhang2021learningisp}. In order to decrease the burden of the Reverse ISP Network, they take advantage of the traditional inverse tone mapping and gamma expansion~\cite{conde2022model, brooks2019unprocessing}.
The pipeline of the method is shown in Figure~\ref{fig:team4}. Note that the team adapts the solution for each track. At Track 1, since the misalignment is small, they use a 9M parameters model (LiteISP $8.88$M, GCM $0.04$M). Considering the spatial misalignment of inputs (RGBs) and targets (RAWs), they adopt a pre-trained PWC-Net~\cite{sun2018pwc} to warp the target RAW images aligned with RGB inputs. GCM~\cite{zhang2021learningisp} is also applied to diminish the effect of color inconsistency in the image alignment. However, since GCM alone is not enough to handle the issue satisfactorily, they use the color histogram to match the GCM output with sRGB images for better color mapping. Based on this framework, they also adopt Spectral Normalization Layer (SN) to improve model's stability and Test-time Local Converter~\cite{chu2021tlc} to alleviate the inconsistency of image sizes between training and testing. By doing this, they can effectively improve performance.
At Track 2, due to the lens shading effect and strong misalignment, they use larger patches of 1536px randomly sampled from the FR image to train the model. Moreover they replace LiteISPNet~\cite{zhang2021learningisp} with NAFNet~\cite{chen2022simplenadnet} as RevISPNet (see Figure~\ref{fig:team4}), this increases model's complexity to 116M parameters.
The method was implemented in PyTorch and trained using 2$\times$ V100 GPU for 5 days using cropped images, AdamW and batch size 6. Flip and rotation $\times$8 data augmentation is used during training. The inference is done using FR images and self-ensembles to improve color inconsistency. The authors emphasize that general FR inference is better than patch-based.

\vspace{-5mm}

\clearpage
\subsection{CS\textasciicircum2U}
\emph{Yue Zhang, Beiji Zou}
\vspace{2mm}

\begin{figure}[!b]
\centering
\includegraphics[width=0.93\textwidth]{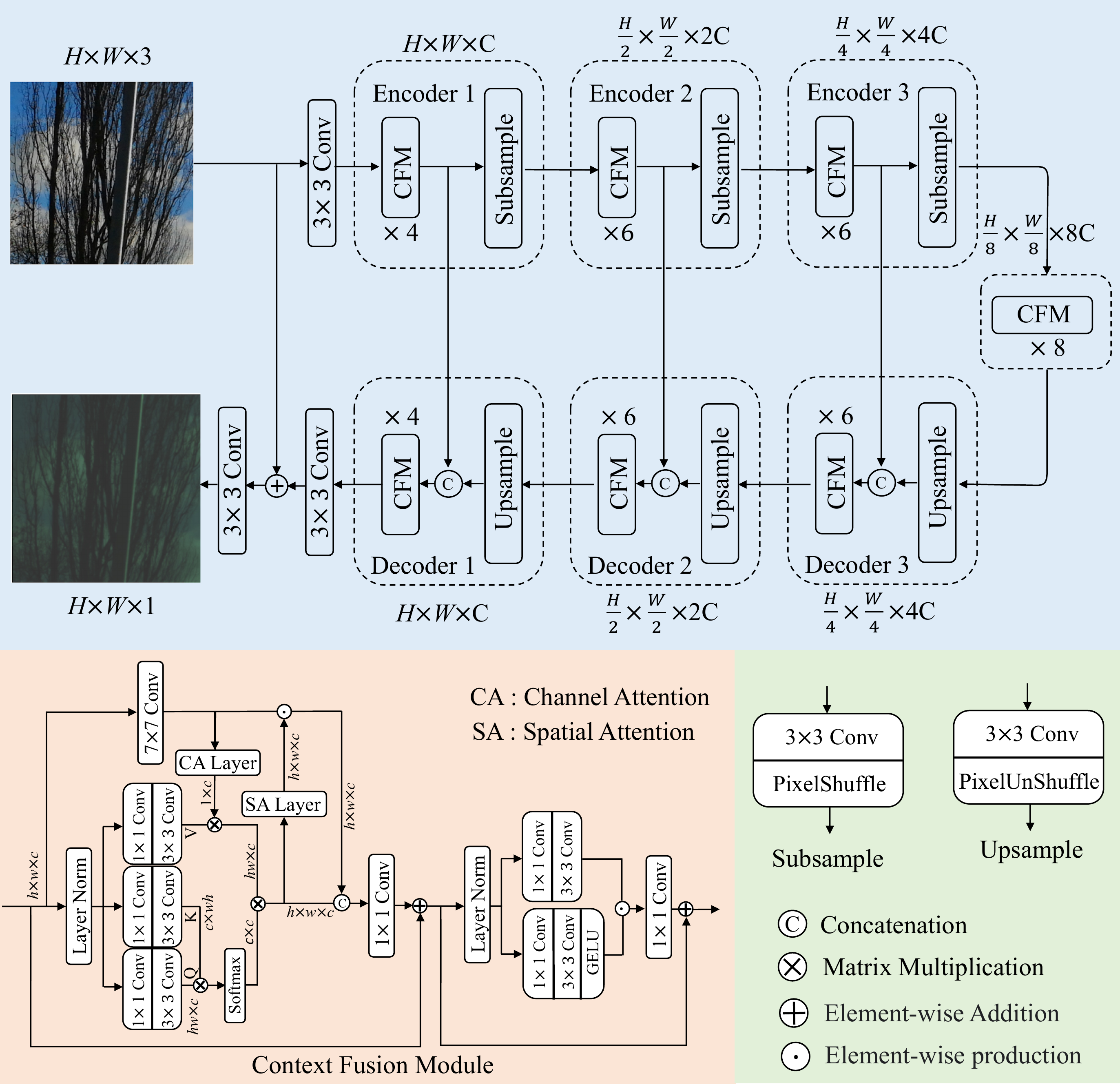}
\caption{SSDNet by CS\textasciicircum2U team.}
\label{fig:team6}
\end{figure}

\noindent The team proposes ``Learned Reverse ISP with soft supervision"~\cite{zou2022learned}, an efficient network and a new loss function for the reversed ISP learning. They propose an encoder-decoder architecture called \textbf{SSDNet} inspired by the Transformer~\cite{chen2022simplenadnet, vaswani2017attention} architecture and is straightforward. The method is illustrated in Figure~\ref{fig:team6}. They exploit the unique property of RAW data, i.e. high dynamic range (HDR), by suggesting to relax the supervision to a multivariate Gaussian distribution in order to learn images that are reasonable for a given supervision. Equipped with the above two components, the method achieves an effective RGB to RAW mapping.
We must note that the team uses the DeepISP S7 dataset~\cite{schwartz2018deepisp}, besides the proposed challenge dataset, this implies that \textbf{the results are not accurate} for this method, since they have potentially trained with extra validation data, nevertheless this should not affect the results in the ``Test2" sets or ``Track2". 


\clearpage
\subsection{SenseBrains}
\emph{Jun Jiang, Jinha Kim}
\vspace{2mm}

\noindent The team proposes ``OEMask-ReverseNet"~\cite{jiang2022overexposure}, a multi-step refinement process integrating an overexposure mask. The key points are: instead of mapping RGB to RAW (Bayer pattern), the pipeline trains from RGB to demosaiced RAW~\cite{gharbi2016deep}; the multi-step process~\cite{brooks2019unprocessing} of this reverse ISP has greatly enhanced the performance of the baseline U-Net; the refinement pipeline can enhance other ``reverse ISPs".
The method is illustrated in Figure~\ref{fig:team7}: (i) unprocess the input RGB~\cite{brooks2019unprocessing} and obtain the demosaiced output~\cite{gharbi2016deep}; (ii) using two independent U-Net~\cite{unet} networks (\texttt{U-Net(OE)}, \texttt{U-Net(NOE)}), estimate an overexposure mask; (iii) apply the mask to the previously obtained demosaiced RGB and finally mosaic it in a 4-channel (RGGB) RAW. They also propose a network to refine the predicted RAWs in the YUV space. The team implemented the method in Pytorch and trained using a combination of LPIPS~\cite{zhang2018lpips}, MS-SSIM-L1 and L2 loss. 

\begin{figure}[]
\centering
\includegraphics[width=0.92\textwidth]{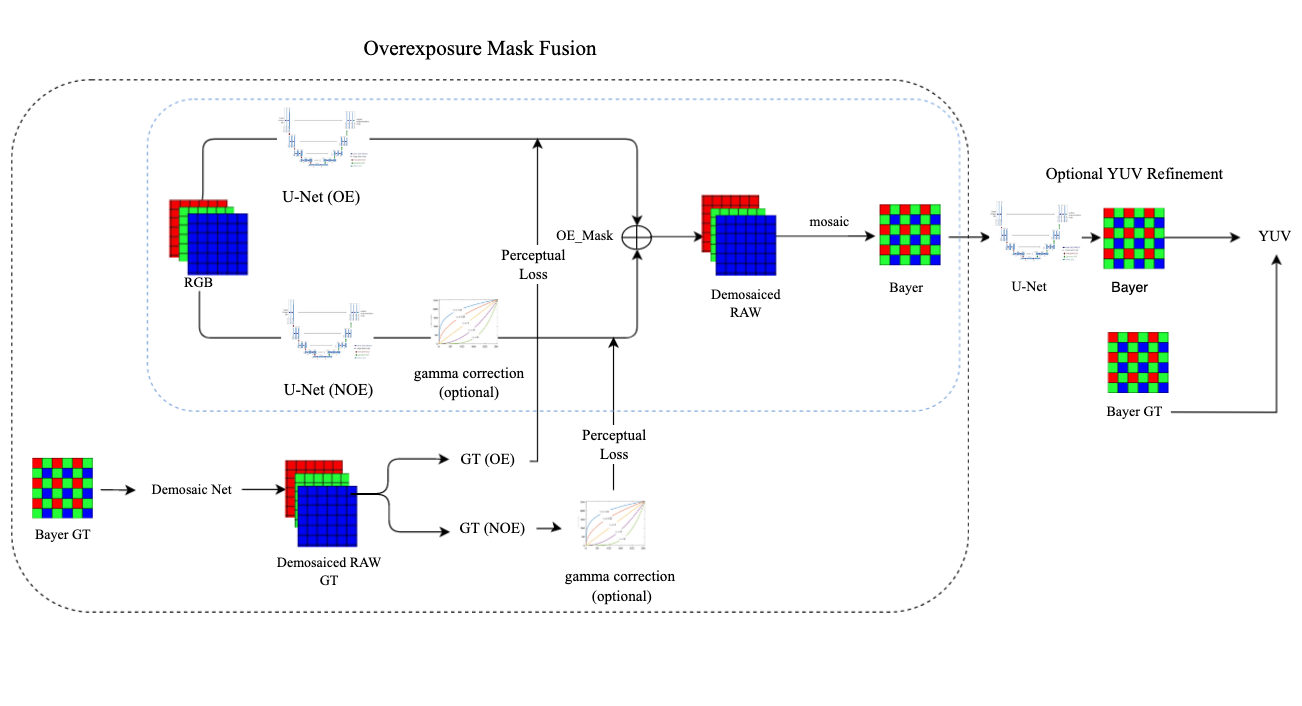}
\caption{``OEMask ReverseNet" by SenseBrains team.}
\label{fig:team5}
\end{figure}

\subsection{HiImage}
\emph{Zhikai Zong, Xiaoxiao Liu}
\vspace{2mm}

\begin{figure}[]
\centering
\includegraphics[width=\textwidth]{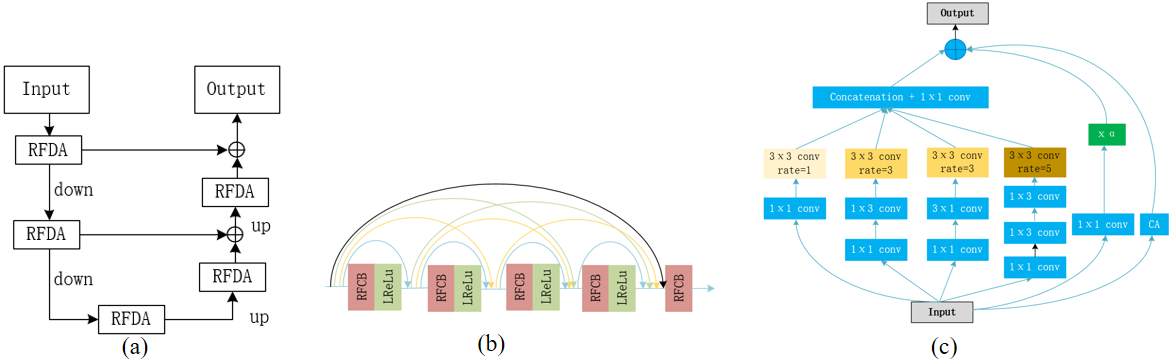}
\caption{``Receptive Field Dense UNet" by HiImage team.}
\label{fig:team7}
\end{figure}

\noindent Team HiImage proposed ``Receptive Field Dense UNet" based on the baseline UNet~\cite{unet}. The method is illustrated in Figure~\ref{fig:team7}. The authors augmented the Receptive Field Dense (RFD) block~\cite{zhang2018residual} with a channel attention module~\cite{woo2018cbam}. In the first stage training, the network was trained to minimize L1 loss using Adam optimizer, learning rate 1e-4, and 256px patches as input. The model has 11M parameters, however, due to the dense connections the number of GFLOPs is high, running time is 200ms per patch in a GPU.

\subsection{0noise}
\emph{Juan Mar{\'i}n Vega, Michael Sloth, Peter Schneider-Kamp, Richard Röttger}
\vspace{2mm}

\begin{figure}[]
\centering
\includegraphics[width=\textwidth]{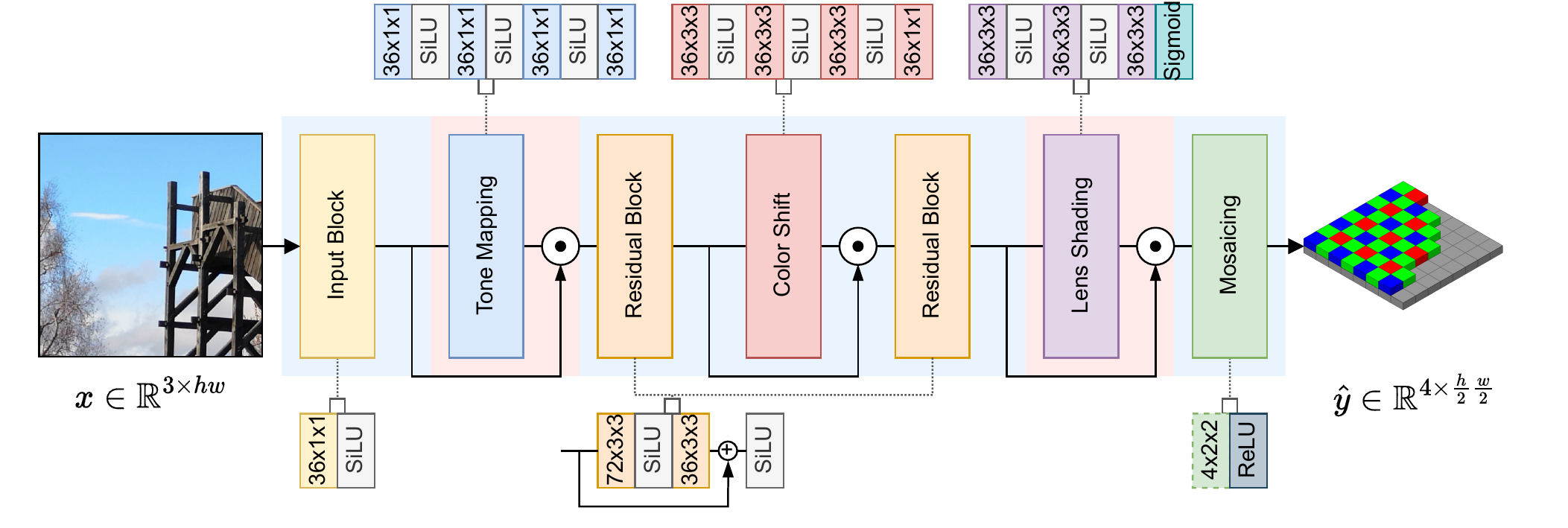}
\caption{``RISP" approach by 0noise team.}
\label{fig:team8}
\end{figure}

\noindent The team introduces RISP, a Reversed ISP network for RGB to RAW mapping. Authors combine model-based~\cite{conde2022model} approaches and custom layers for an end-to-end RGB o RAW mapping network. In particular, they utilize Tone Mapping and Lens Correction layers from MBISPLD~\cite{conde2022model}, and combine it with a custom Color Shift and a Mosaicing Layer. The method is illustrated in Figure~\ref{fig:team8}. The tone mapping, color shift and lens shading are applied by a point-wise multiplication. The mosaicing layer produces an output of $\frac{h}{2}\frac{w}{2}$ with a 2x2 convolution with stride 2.
The method is implemented using Pytorch. The network was trained using Adam for 200 epochs and learning rate $1e-4$, and $250\times250$ patches. Authors use a combination of the L1 and VGG19 losses.
We must highlight that this model is the \textbf{smallest among the proposed ones}, the model has only 0.17M parameters, it is fast and able to process 4K images under a second.
Based on internal experiments, authors emphasize that there is no clear benefit in using a computationally heavy network as CycleISP~\cite{zamir2020cycleisp} in comparison to lighter or model-based approaches~\cite{conde2022model}. The code will be open-sourced.

\subsection{OzU VGL}
\emph{Furkan Kınlı, Barış Özcan, Furkan Kıraç}
\vspace{2mm}

\begin{figure}[!ht]
\centering
\includegraphics[width=0.94\textwidth]{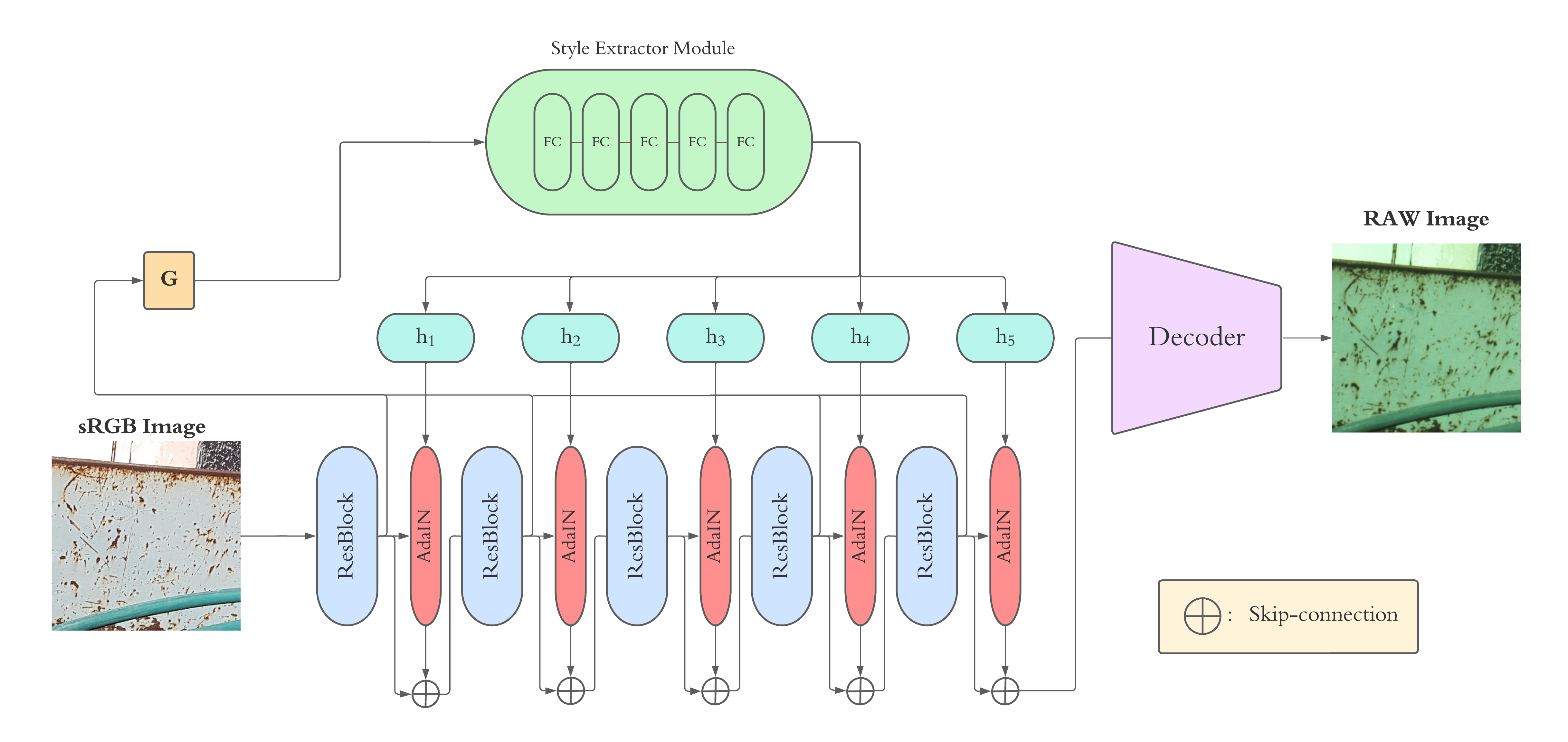}
\caption{``Reversed ISP by Reverse Style Transfer" approach by OzU VGL.}
\label{fig:team9}
\end{figure}

\noindent The team proposes ``Reversed ISP by Reverse Style Transfer"~\cite{kinli2022reversing}. Based on their previous work~\cite{Kinli_2021_CVPR, Kinli_2022_CVPR}, they propose to model the non-linear operations required to reverse the ISP pipeline as the style factor by using reverse style transferring strategy. The architecture has an encoder-decoder structure as illustrated in Figure~\ref{fig:team9}. Authors assume the final sRGB output image has additional injected style information on top of the RAW image, and this additional information can be removed by adaptive feature normalization. Moreover, they employ a wavelet-based discriminator network, which provides a discriminative regularization to the final RAW output of the main network. This architecture has 86.3M parameters, mostly due to the style projectors for each residual block. This method was implemented in PyTorch and trained for 52 epochs using patches, Adam with the learning rate of $1e-4$. The code will be open-sourced.

\vspace{-5mm}

\vspace{-4mm}

\subsection{PixelJump}
\emph{Li Leyi}
\vspace{2mm}

\noindent The proposed solution was focused on aligned image pairs (Track 1). The mapping between RAW and sRGB is divided into two parts: pixel-independent mapping (\eg{} white balance, color space transformation) and local-dependent mapping (\eg{} local engancement). 
The proposed method implemented the \textbf{pixel-level} mapping as ``backend" using global colour mapping module proposed in~\cite{chen2021new} to learn the pixel-independent mapping between RAW and sRGB. Then, the \textbf{local enhancement} module (encoder-decoder~\cite{zhu2020eednet} and ResNet structure~\cite{he2016identity}) is added to deal with the local-dependent mapping.
Ultimately, the outputs of these three modules were fused by a set of learned weights from the weight predictor inspired by~\cite{zeng2020learning}.
The method is implemented in PyTorch and trained using 8$\times$ RTX 3090 GPU, and batch size 2; it processes a patch in 40ms on GPU.

\begin{figure}[]
\centering
\includegraphics[width=\textwidth, keepaspectratio]{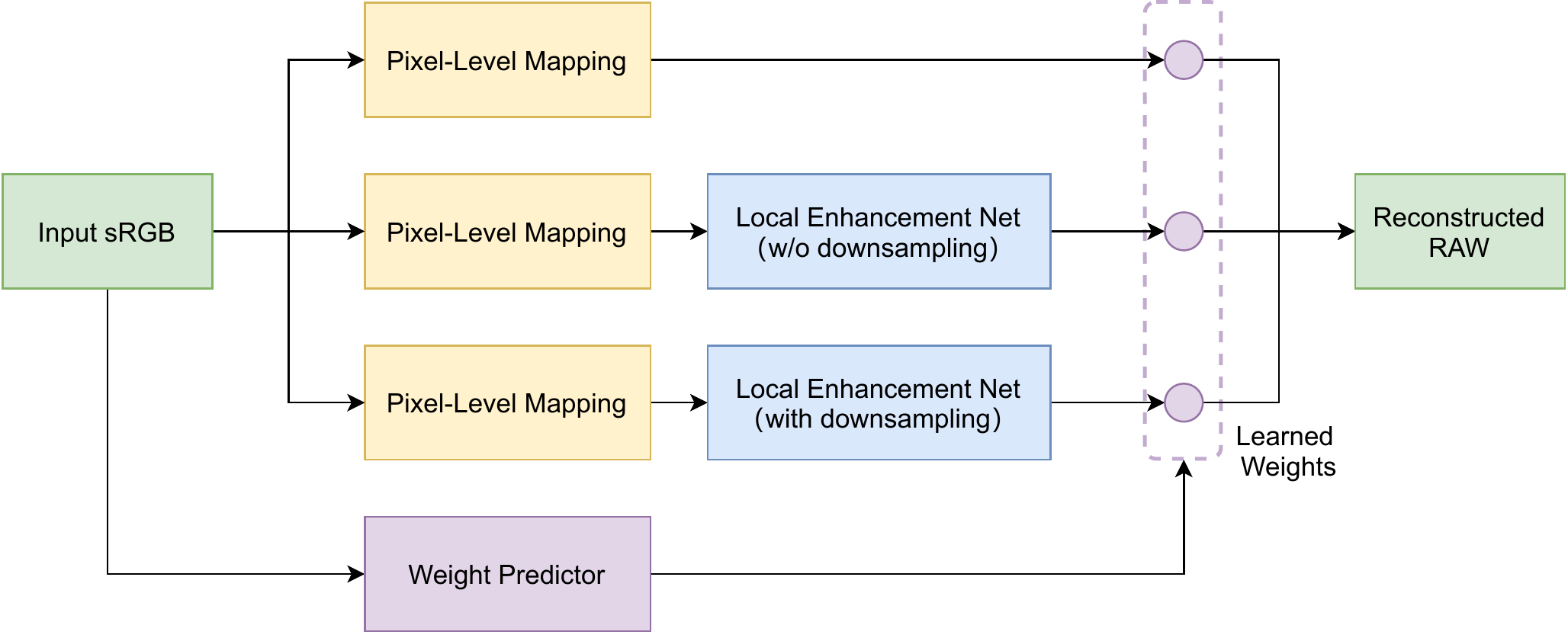} 
\caption{PixelJump method overview.}
\label{fig:team10}
\end{figure}

\subsection{CVIP}
\emph{SM Nadim Uddin, Dipon Kumar Ghosh, Yong Ju Jung}
\vspace{2mm}

\begin{figure}[]
\centering
\includegraphics[width=\textwidth, keepaspectratio]{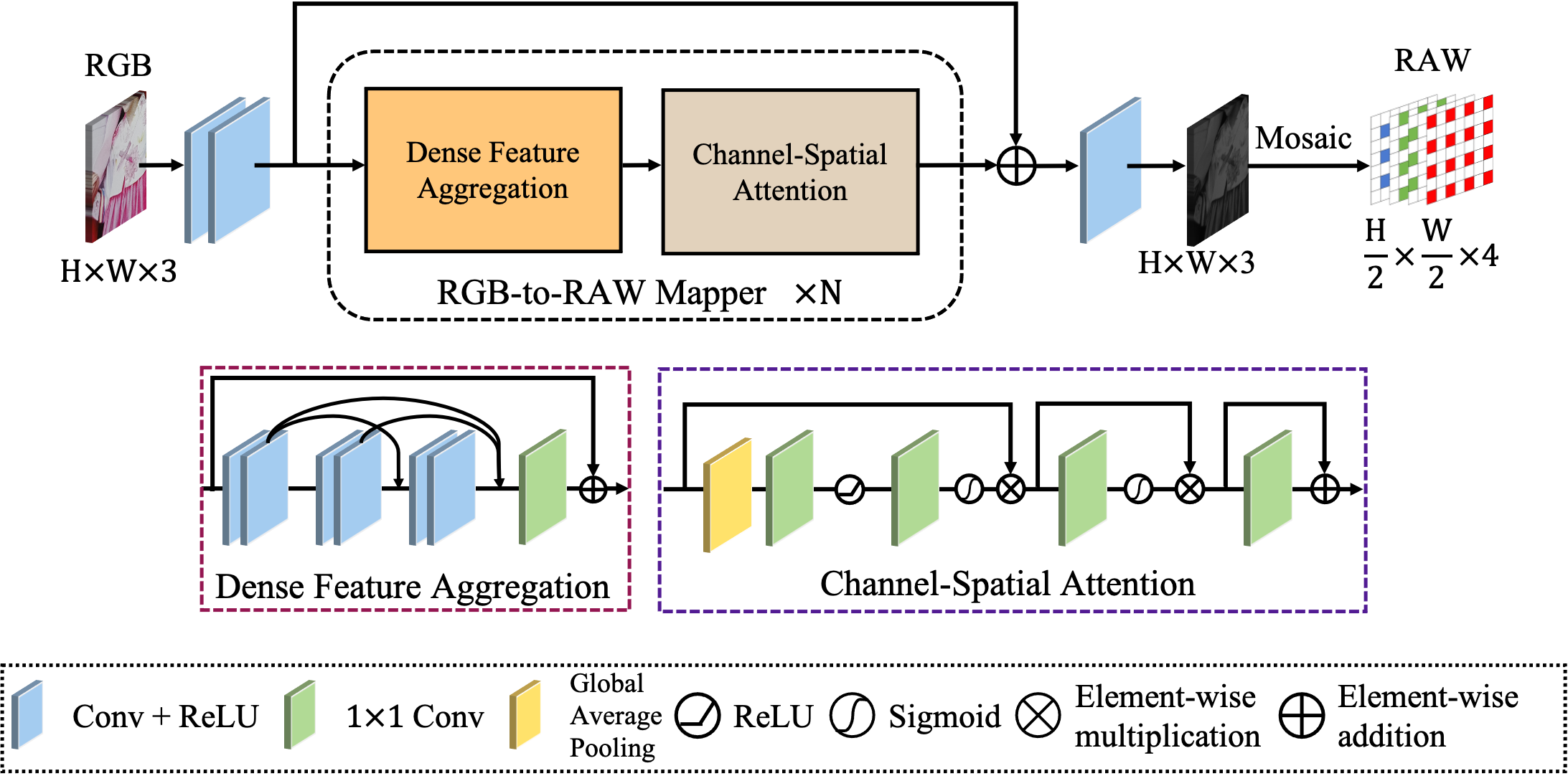} 
\caption{CVIP method overview.}
\label{fig:team11}
\end{figure}

\noindent The team proposes ``Reverse ISP Estimation using Recurrent Dense Channel-Spatial Attention". This method is illustrated in Figure~\ref{fig:team11}. Authors compare their method directly with CycleISP~\cite{zamir2020cycleisp} and improve considerably (+5dB).
The method was implemented in PyTorch 1.8 and trained for 2 days using a RTX 3090 GPU. MAD and SSIM were used as loss functions. The method has only 2.8M parameters (724 GFLOP) and it can unprocess a $252$px RGB in 10ms (using GPU).
This method can generalize to unseen sensors as shown in Table~\ref{tab:bench}, however, authors detected a bias towards indoor images.

\vspace{-1mm}

\begin{table}[!hb]
    \centering
    \caption{Team information summary. ``Input" refers to the input image size used during training, most teams used the provided patches (\ie 504px or 496px). ED indicates the use of ``Extra Datasets" besides the provided challenge datasets. ENS indicates if the solution is an ``Ensemble" of multiple models. FR indicates if the model can process ``Full-Resolution" images (\ie $3024\times4032$)}
    \vspace{2mm}
    \label{tab:teams}
    \resizebox{\linewidth}{!}{
    \begin{tabular}{l|c|c|c|c|c|c|c|c}
        \hline
        Team & Input & Epochs & ED & ENS & FR  & \# Params. (M) & Runtime (ms) & GPU \\
        \hline
        \rowcolor{Gray} NOAHTCV & (504,504) & 500 & \xmark & \xmark & \cmark & 5.6 & 25 & V100 \\
        MiAlgo & (3024,4032) & 3000 & \xmark & \xmark & \cmark & 4.5 & 18 & V100 \\
        \rowcolor{Gray} CASIA LCVG & (504,504) & 300K it. & \cmark & \cmark & \cmark & 464 & 219 & A100 \\
        CS\textasciicircum2U & (504,504) & 276K it. & \cmark & \cmark & \cmark & 105 & 1300 & 3090 \\
        \rowcolor{Gray} HIT-IIL & (1536,1536) & 1000 & \xmark & \xmark & \cmark & 9/116 & 19818 (cpu) & V100 \\
        SenseBrains & (504,504) & 220 & \xmark & \cmark & \cmark & 69 & 50 & V100 \\
        \rowcolor{Gray} PixelJump & (504,504) & 400  & \xmark & \cmark & \cmark & 6.64 & 40 & 3090 \\
        HiImage & (256, 256) & 600 & \xmark & \xmark & \cmark & 11 & 200 & 3090 \\
        \rowcolor{Gray} OzU VGL & (496, 496) & 52 & \xmark & \xmark & \cmark & 86 & 6 & 2080 \\
        CVIP & (504,504) & 75 & \xmark & \xmark & \cmark & 2.8 & 400 & 3090 \\
        \rowcolor{Gray} 0noise & (504,504) & 200 & \xmark & \xmark & \cmark & 0.17 & 19 & Q6000 \\
        \hline
    \end{tabular}
    }
\end{table}

\vspace{-3.5mm}
\section{Conclusions}

The AIM 2022 Reversed ISP Challenge promotes a novel direction of research aiming at solving the availability of RAW image datasets by leveraging the abundant and diverse RGB counterparts. To the best of our knowledge, this is the first work that analyzes how to recover raw sensor data from the processed RGBs (\ie reverse the ISP transformations) using real smartphones and learned methods, and evaluating the quality of the unprocessed synthetic RAW images.
The proposed task is essentially an ill-posed inverse problem, learning an approach to approximate inverse functions of real-world ISPs have implicit \textbf{limitations}, for instance the quantization of the 14-bit RAW image to the 8-bit RGB image lead to inevitable information lost.
The presented methods and benchmark gauge the state-of-the-art in RAW image reconstruction.
As we can see in Table~\ref{tab:teams}, most methods can unprocess full-resolution RGBs under 1s on GPU.

We believe this problem and the proposed solutions can have a positive impact in multiple low-level vision tasks that operate in the RAW domain such as denoising, demosaicing, hdr or super-resolution.
For \textbf{reproducibility} purposes, we provide more information and the code of many top solutions at:\\
\url{https://github.com/mv-lab/AISP/tree/main/aim22-reverseisp}

\section*{Acknowledgments}
This work was supported by the Humboldt Foundation.
We thank the sponsors of the AIM and Mobile AI 2022 workshops and challenges: AI Witchlabs, MediaTek, Huawei, Reality Labs, OPPO, Synaptics, Raspberry Pi, ETH Z\"urich (Computer Vision Lab) and University of W\"urzburg (Computer Vision Lab).
We also thank Andrey Ignatov and Eli Schwartz for their datasets used in this challenge.

\clearpage
%
%

{\small
\bibliographystyle{splncs04}
\bibliography{egbib}
}


\clearpage

\appendix

\section{Appendix 1: Qualitative Results}
\label{app:extra-quali-results}

As we pointed out in Section~\ref{sec:eval}, we find a clear fidelity-perception tradeoff~\cite{blau2018perception} in the solutions, \ie{} some methods achieve high PSNR values, yet the generated RAW images do no look realistic.
In this report, due to the visualization and space constraints, we only include samples from the best 8 ranked teams in Table~\ref{tab:bench}.
We provide dditional qualitative comparisons including all the teams, and supplementary material at:\\ \url{https://github.com/mv-lab/AISP/tree/main/aim22-reverseisp}

\begin{figure}[ht]
    \centering
    \setlength{\tabcolsep}{2.0pt}
    \begin{tabular}{cccc}
    \includegraphics[width=0.24\linewidth]{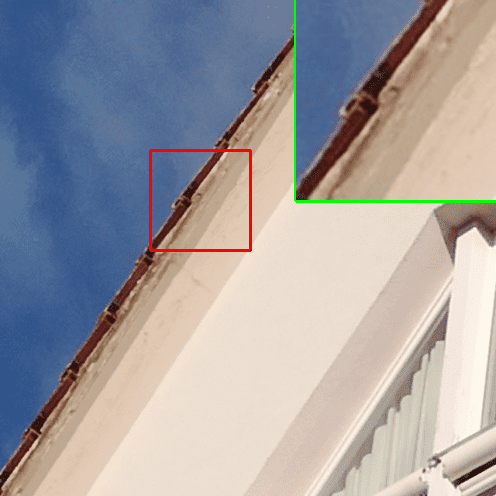} &
    \includegraphics[width=0.24\linewidth]{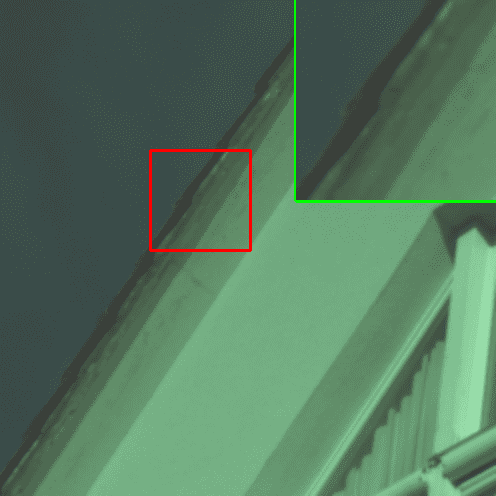} &
    \includegraphics[width=0.24\linewidth]{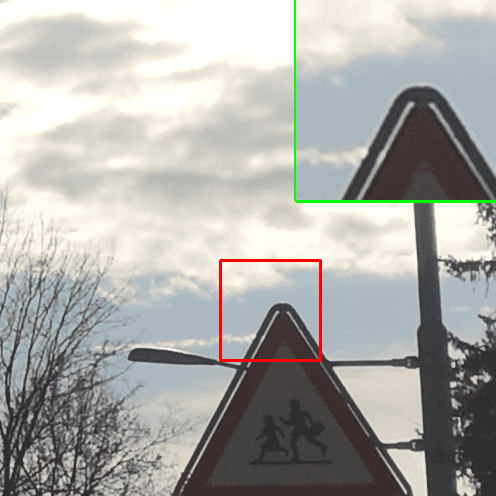} &
    \includegraphics[width=0.24\linewidth]{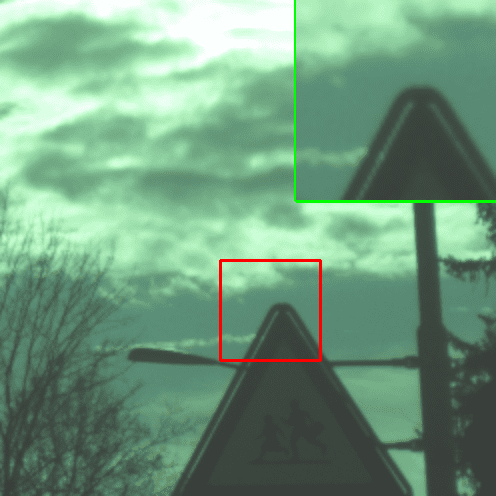}
    \tabularnewline
    \multicolumn{2}{c}{Ground-truth RGB-RAW Scene 3} & \multicolumn{2}{c}{Ground-truth RGB-RAW Scene 4}
    \vspace{3mm}
    \tabularnewline
    \includegraphics[width=0.24\linewidth]{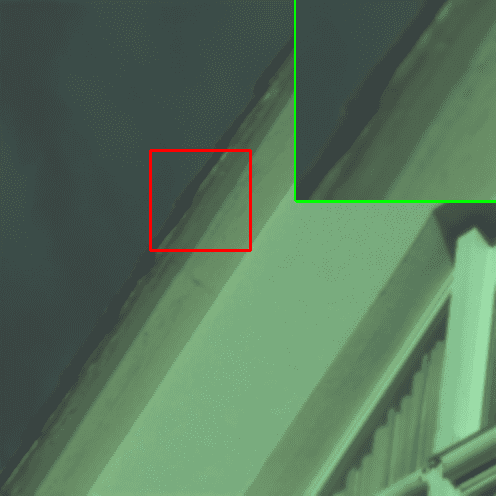} &
    \includegraphics[width=0.24\linewidth]{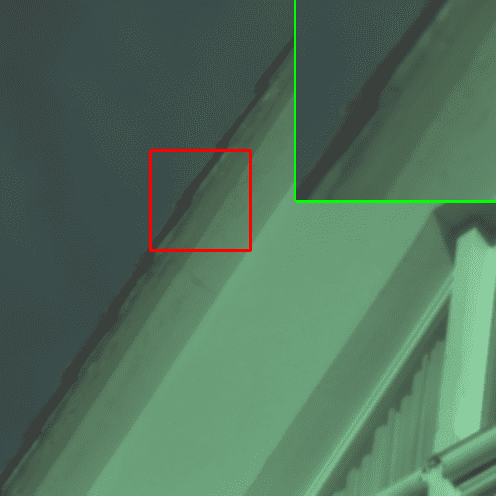} &
    \includegraphics[width=0.24\linewidth]{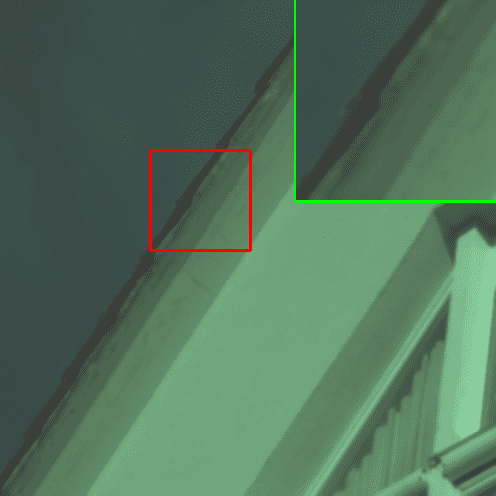} &
    \includegraphics[width=0.24\linewidth]{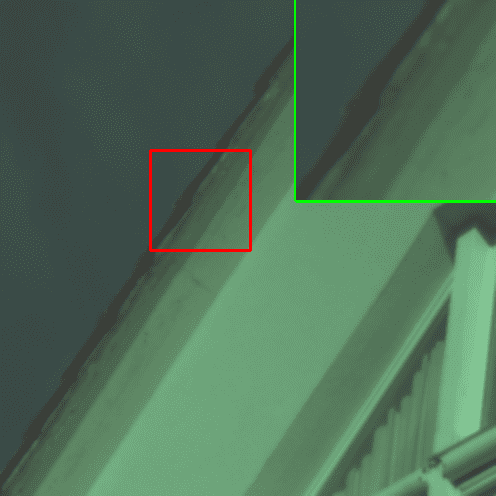}
    \tabularnewline
    NOAHTCV & MiAlgo & CASIA LCVG & HIT-IIL
    \tabularnewline
    \includegraphics[width=0.24\linewidth]{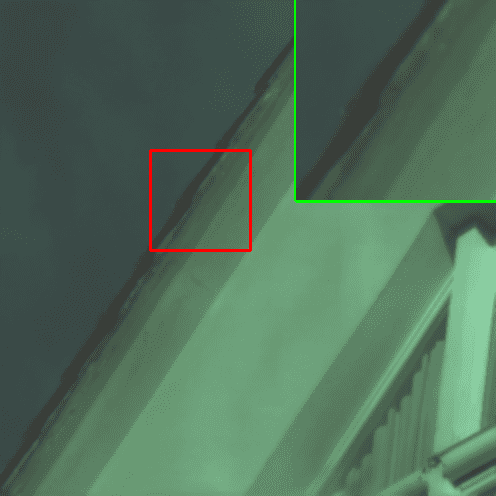} &
    \includegraphics[width=0.24\linewidth]{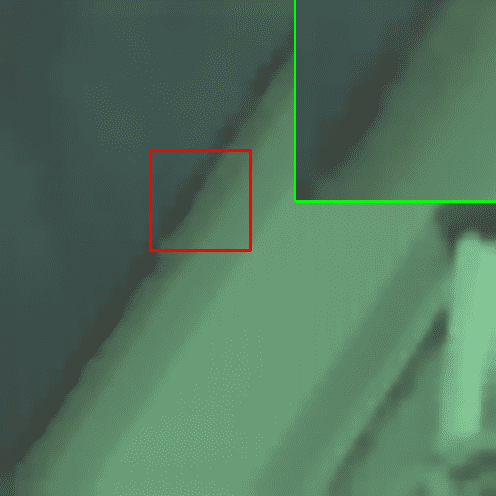} &
    \includegraphics[width=0.24\linewidth]{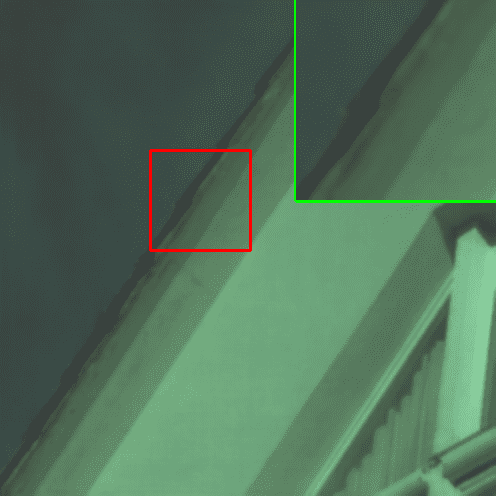} &
    \includegraphics[width=0.24\linewidth]{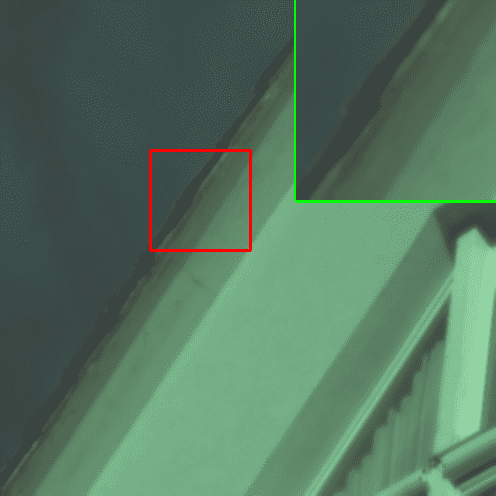}
    \tabularnewline
    SenseBrains & OzU VGL & HiImage & 0noise
    \vspace{10mm}
    \tabularnewline
    \includegraphics[width=0.24\linewidth]{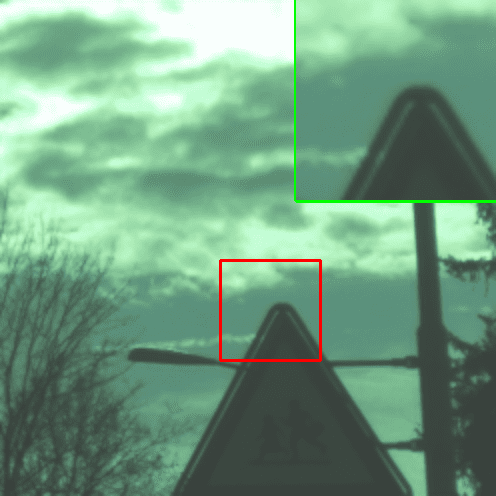} &
    \includegraphics[width=0.24\linewidth]{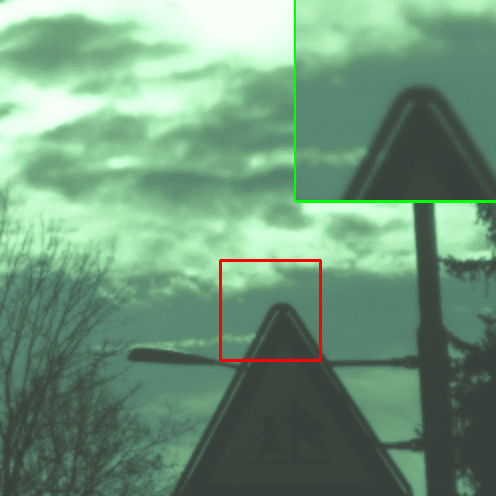} &
    \includegraphics[width=0.24\linewidth]{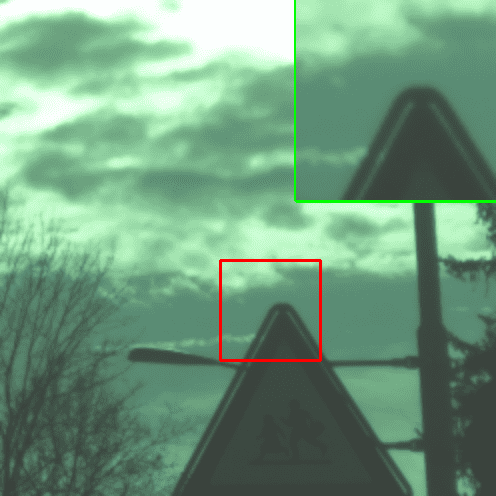} &
    \includegraphics[width=0.24\linewidth]{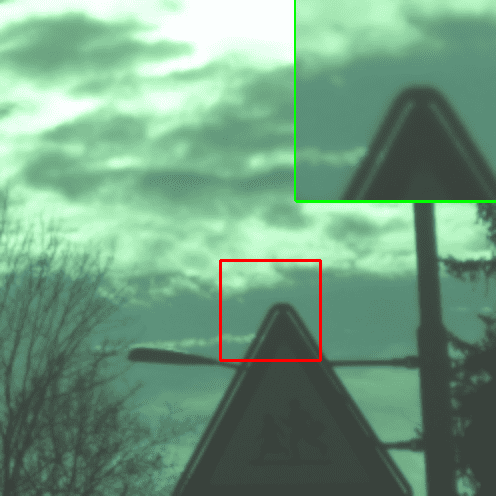}
    \tabularnewline
    NOAHTCV & MiAlgo & CASIA LCVG & HIT-IIL
    \tabularnewline
    \includegraphics[width=0.24\linewidth]{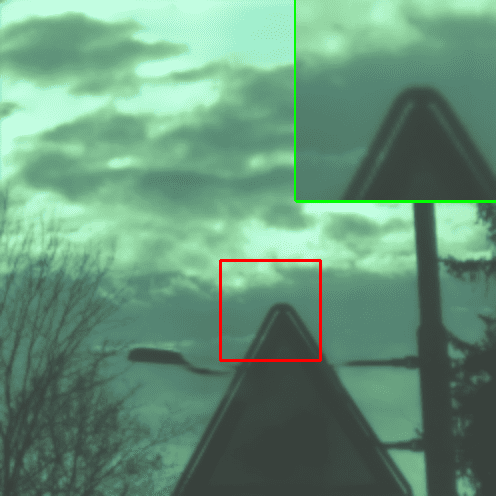} &
    \includegraphics[width=0.24\linewidth]{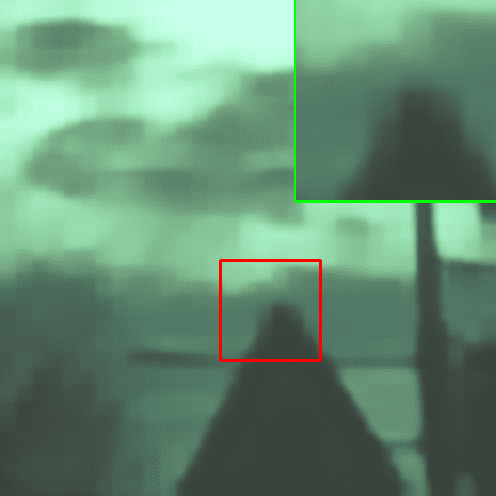} &
    \includegraphics[width=0.24\linewidth]{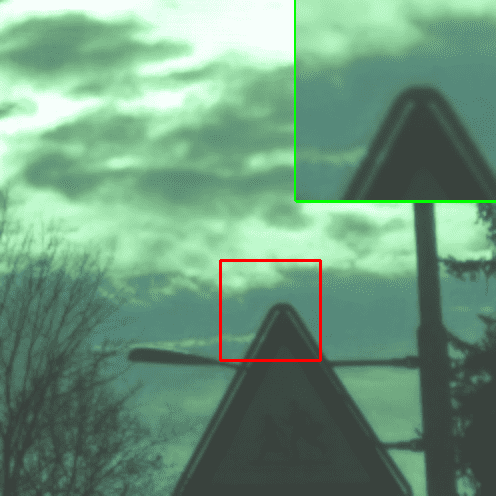} &
    \includegraphics[width=0.24\linewidth]{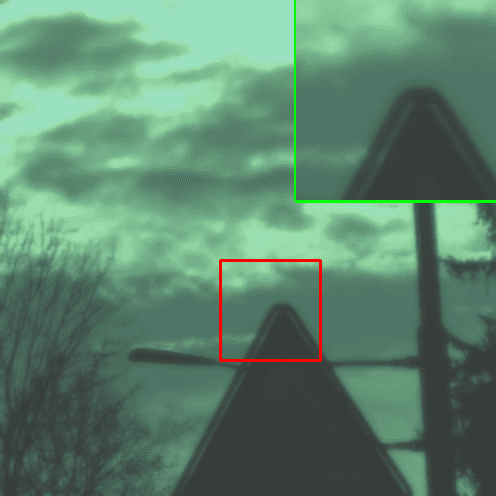}
    \tabularnewline
    SenseBrains & OzU VGL & HiImage & 0noise
    \tabularnewline
    \end{tabular}
    \caption{Qualitative Comparison on the Track 2 (P20). Details and intensities are recovered even for these non-aligned images (\ie slightly overexposed sky).}
    \label{fig:quali-results3}
\end{figure}

\begin{figure}[ht]
    \centering
    \setlength{\tabcolsep}{2.0pt}
    \begin{tabular}{cccc}
    \includegraphics[width=0.24\linewidth]{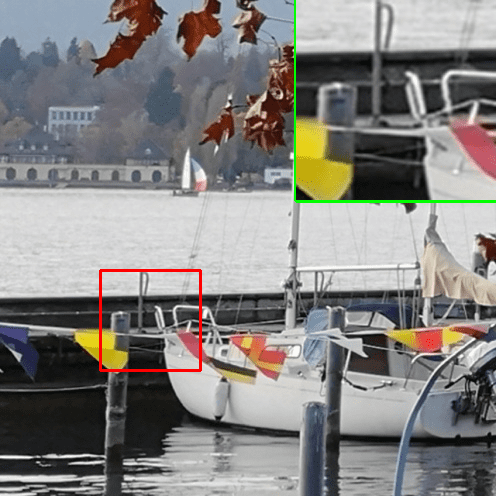} &
    \includegraphics[width=0.24\linewidth]{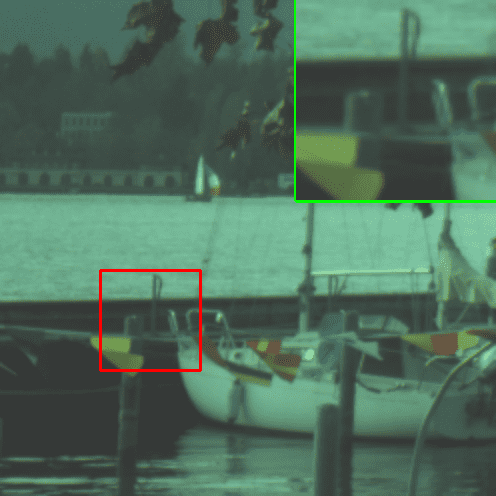} &
    \includegraphics[width=0.24\linewidth]{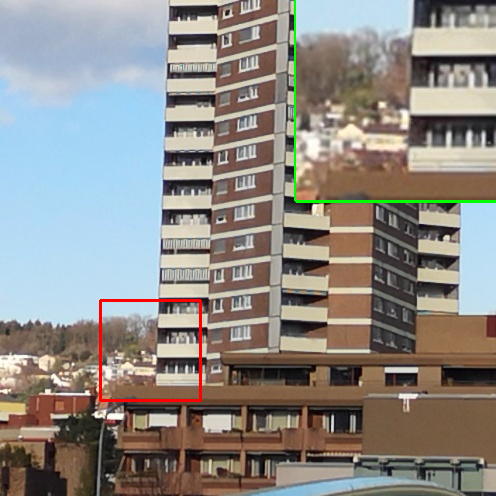} &
    \includegraphics[width=0.24\linewidth]{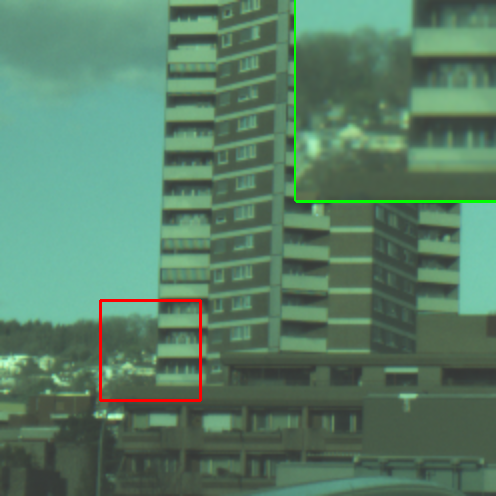}
    \tabularnewline
    \multicolumn{2}{c}{Ground-truth RGB-RAW Scene 5} & \multicolumn{2}{c}{Ground-truth RGB-RAW Scene 6}
    \vspace{3mm}
    \tabularnewline
    \includegraphics[width=0.24\linewidth]{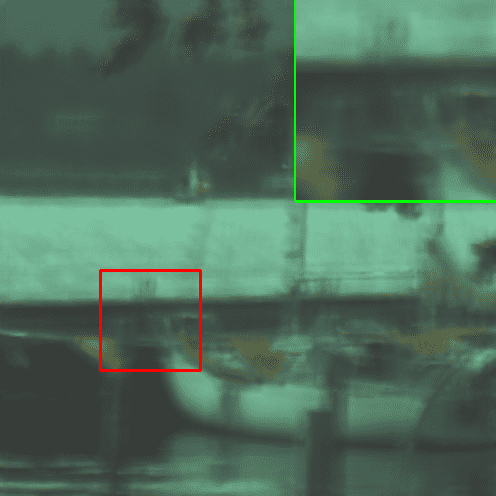} &
    \includegraphics[width=0.24\linewidth]{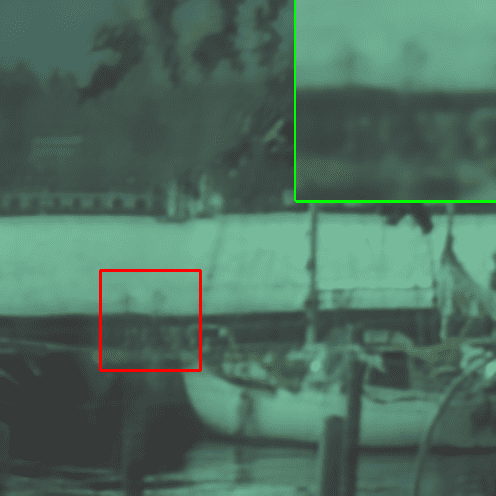} &
    \includegraphics[width=0.24\linewidth]{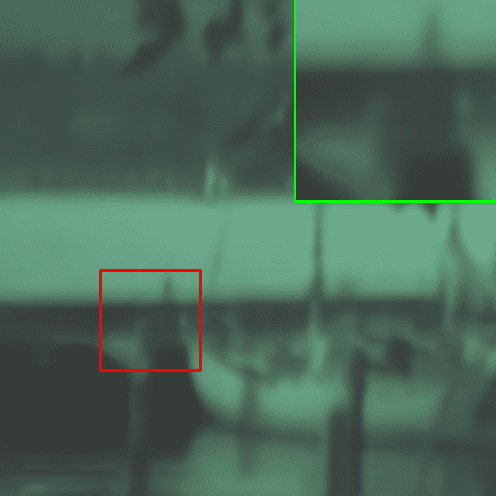} &
    \includegraphics[width=0.24\linewidth]{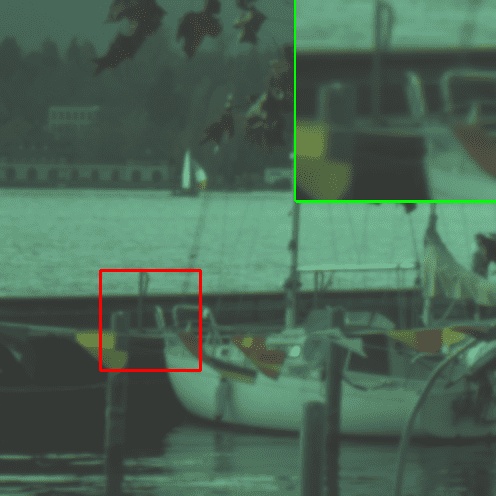}
    \tabularnewline
    NOAHTCV & MiAlgo & CASIA LCVG & HIT-IIL
    \tabularnewline
    \includegraphics[width=0.24\linewidth]{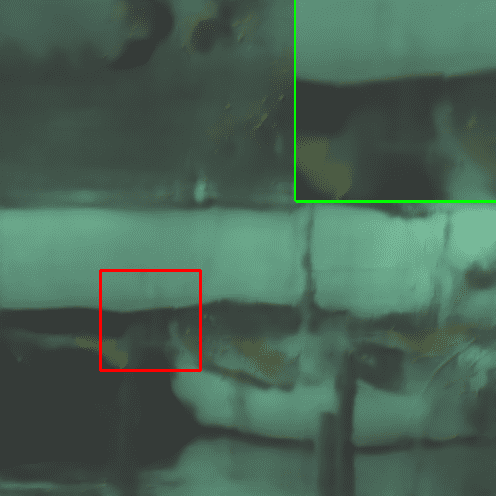} &
    \includegraphics[width=0.24\linewidth]{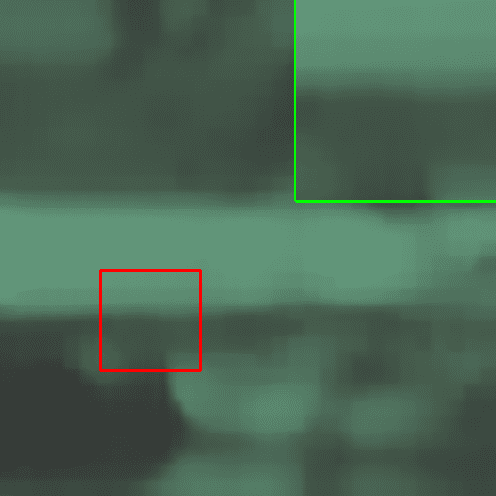} &
    \includegraphics[width=0.24\linewidth]{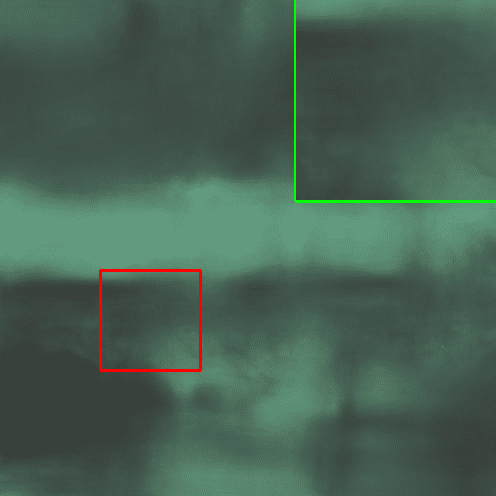} &
    \includegraphics[width=0.24\linewidth]{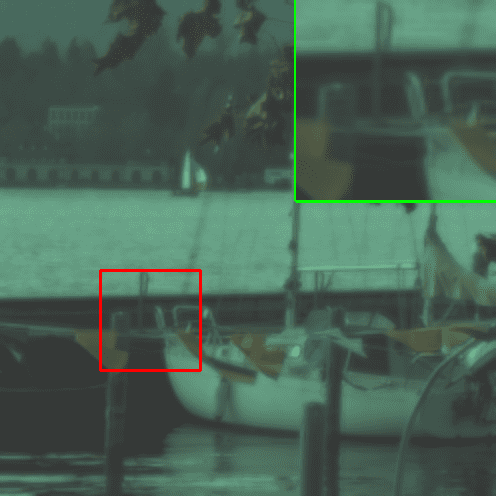}
    \tabularnewline
    SenseBrains & OzU VGL & HiImage & 0noise
    \vspace{10mm}
    \tabularnewline
    \includegraphics[width=0.24\linewidth]{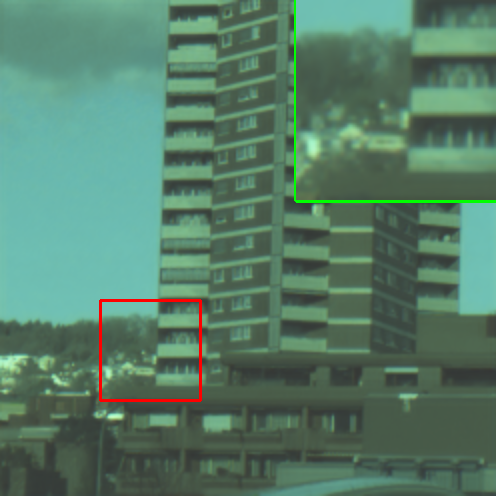} &
    \includegraphics[width=0.24\linewidth]{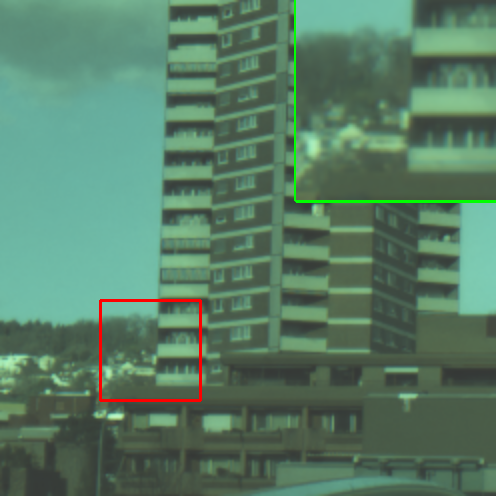} &
    \includegraphics[width=0.24\linewidth]{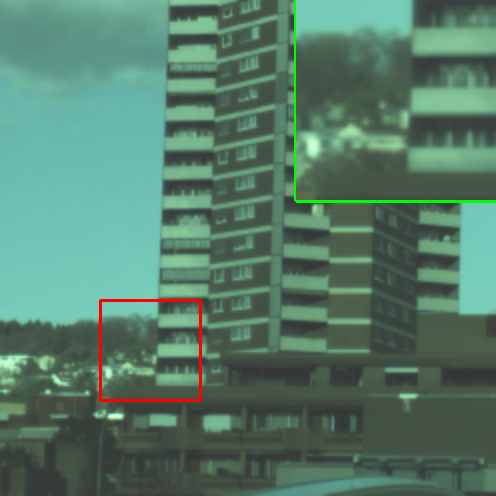} &
    \includegraphics[width=0.24\linewidth]{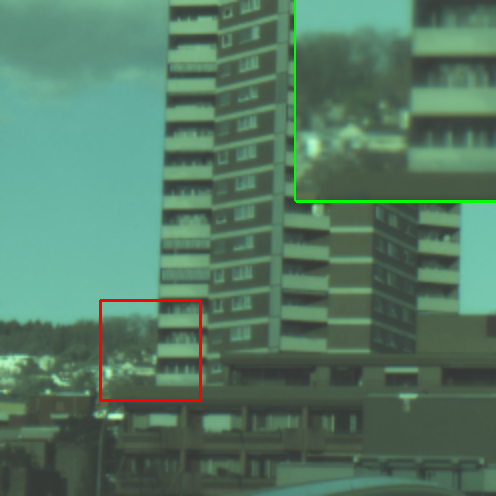}
    \tabularnewline
    NOAHTCV & MiAlgo & CASIA LCVG & HIT-IIL
    \tabularnewline
    \includegraphics[width=0.24\linewidth]{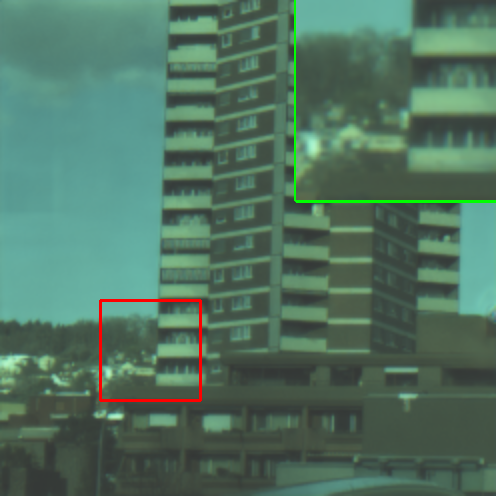} &
    \includegraphics[width=0.24\linewidth]{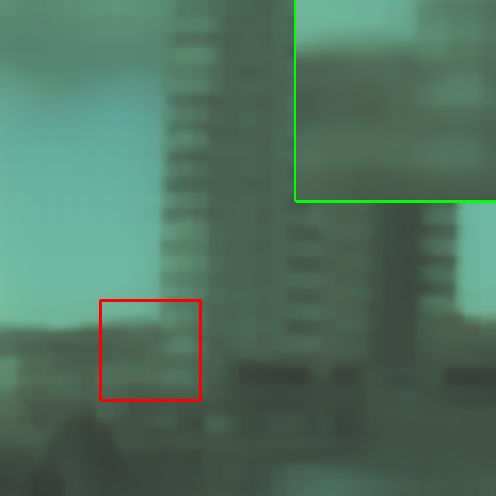} &
    \includegraphics[width=0.24\linewidth]{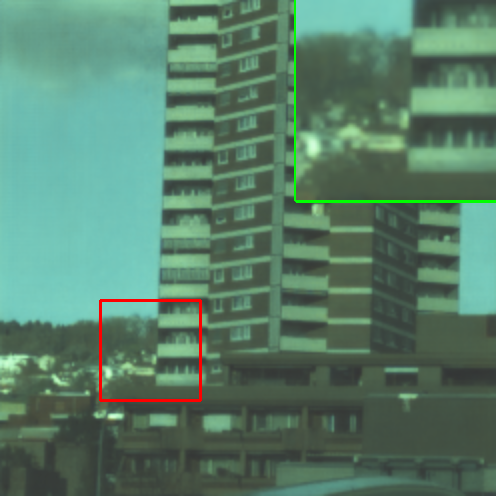} &
    \includegraphics[width=0.24\linewidth]{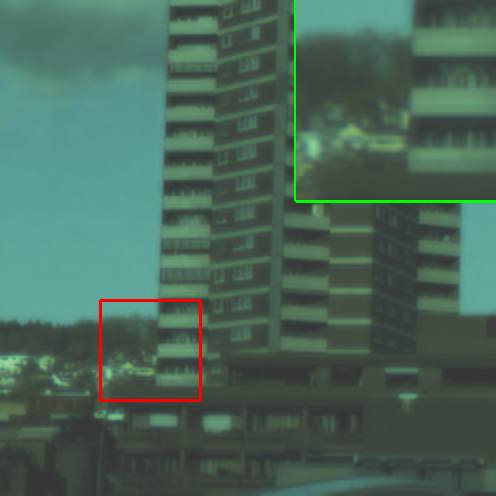}
    \tabularnewline
    SenseBrains & OzU VGL & HiImage & 0noise
    \tabularnewline
    \end{tabular}
    \caption{Qualitative Comparison of the top performing methods on Track 2 (P20). In these challenging non-aligned samples we can appreciate clear differences between methods, many cannot recover high-frequencies and produce blur artifacts.}
    \label{fig:quali-results4}
\end{figure}

The following figures~\ref{fig:quali-results1}~\ref{fig:quali-results2}~\ref{fig:quali-results3}~\ref{fig:quali-results4} show a complete comparison between the most competitive methods in Track 1 and 2. Overall, the RAW reconstruction quality is high, however, we can appreciate clear colour and texture differences between the methods. Note that the RAW images are visualized using a simple green-average demosaic, tone mapping and gamma operators. Moreover, in the case of the Track 2 (P20) images are strongly misaligned.

\begin{figure}[ht]
    \centering
    \setlength{\tabcolsep}{2.0pt}
    \begin{tabular}{cccc}
    \multicolumn{2}{c}{\includegraphics[width=0.25\linewidth]{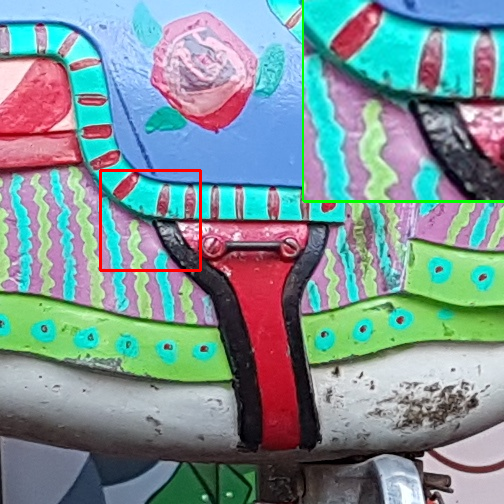}} &
    \multicolumn{2}{c}{\includegraphics[width=0.25\linewidth]{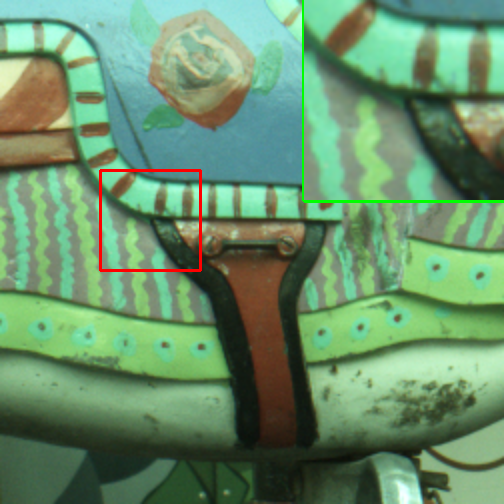}}
    \tabularnewline
    \multicolumn{2}{c}{Ground-truth RGB}&
    \multicolumn{2}{c}{Ground-truth RAW}
    \vspace{3mm}
    \tabularnewline
    \includegraphics[width=0.24\linewidth]{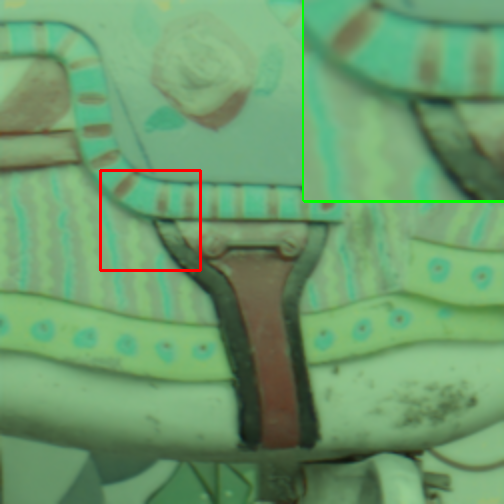} &
    \includegraphics[width=0.24\linewidth]{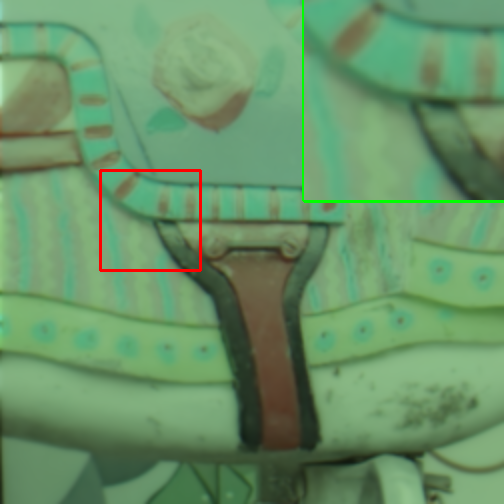} &
    \includegraphics[width=0.24\linewidth]{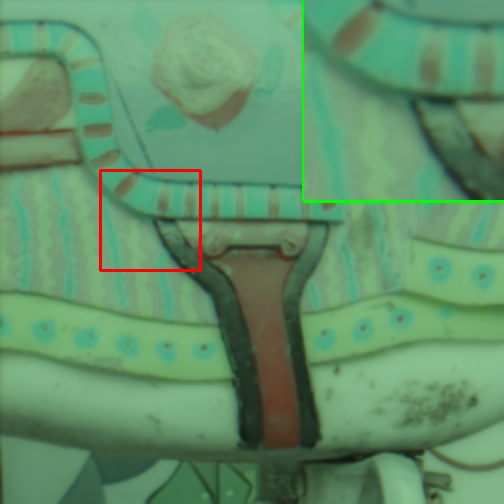} &
    \includegraphics[width=0.24\linewidth]{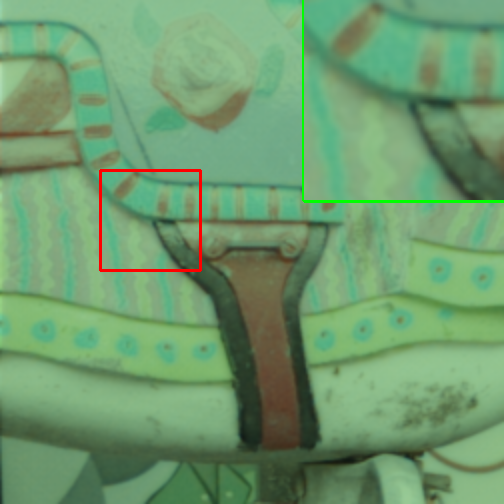}
    \tabularnewline
    NOAHTCV & MiAlgo & CASIA LCVG & HIT-IIL 
    \tabularnewline
    \includegraphics[width=0.24\linewidth]{figures/samples/t1/48_19_NOAHTCV.png} &
    \includegraphics[width=0.24\linewidth]{figures/samples/t1/48_19_MiAlgo.png} &
    \includegraphics[width=0.24\linewidth]{figures/samples/t1/48_19_CASIA.png} &
    \includegraphics[width=0.24\linewidth]{figures/samples/t1/48_19_HIT-IIL.png}
    \tabularnewline
    SenseBrains & CSˆ2U & HiImage & 0noise
    \tabularnewline
    \end{tabular}
    \caption{Qualitative Comparison on the Track 1 (S7). We can appreciate that most methods recover accurately colors, yet struggle at recovering light intensity.}
    \label{fig:quali-results1}
\end{figure}

\begin{figure}[ht]
    \centering
    \setlength{\tabcolsep}{2.0pt}
    \begin{tabular}{cccc}
    \includegraphics[width=0.24\linewidth]{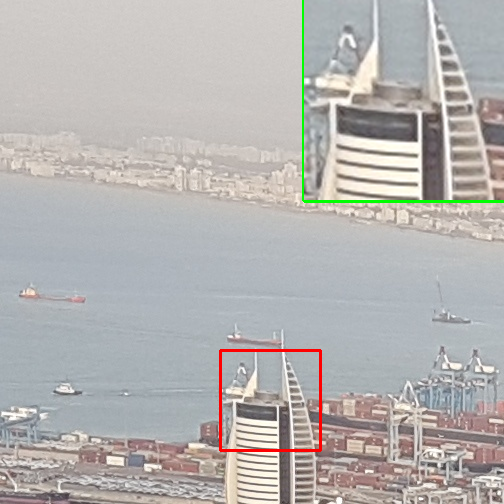} &
    \includegraphics[width=0.24\linewidth]{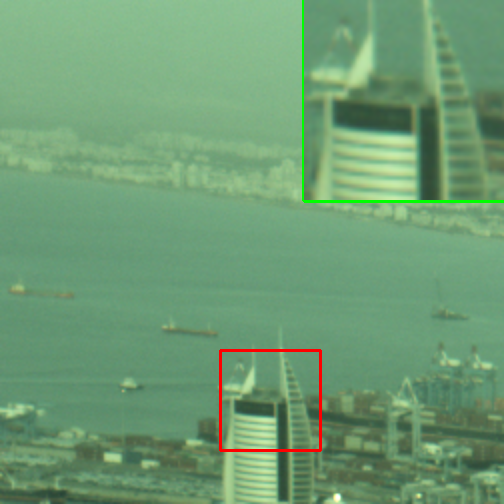} &
    \includegraphics[width=0.24\linewidth]{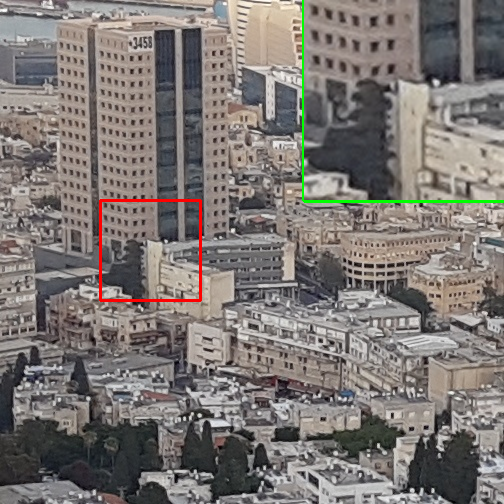} &
    \includegraphics[width=0.24\linewidth]{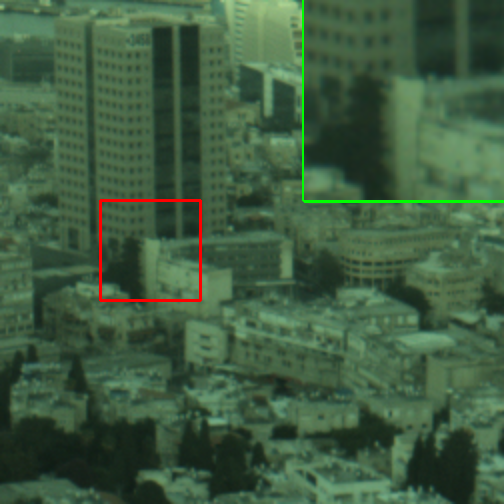}
    \tabularnewline
    \multicolumn{2}{c}{Ground-truth RGB-RAW Scene 1} & \multicolumn{2}{c}{Ground-truth RGB-RAW Scene 2}
    \vspace{3mm}
    \tabularnewline
    \includegraphics[width=0.24\linewidth]{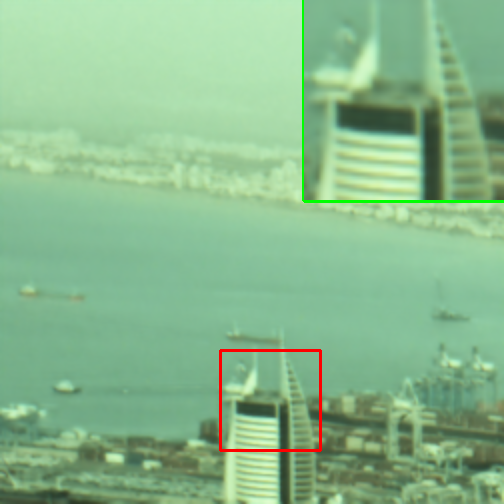} &
    \includegraphics[width=0.24\linewidth]{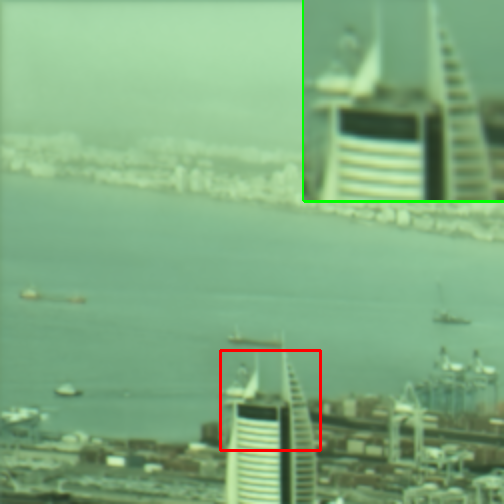} &
    \includegraphics[width=0.24\linewidth]{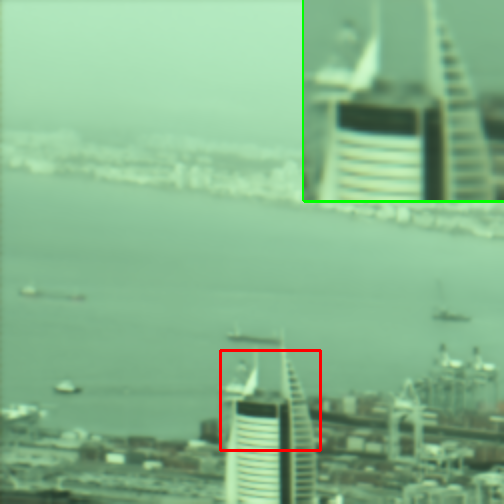} &
    \includegraphics[width=0.24\linewidth]{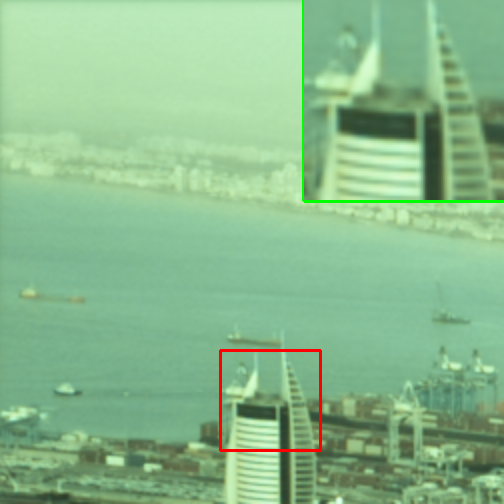}
    \tabularnewline
    NOAHTCV & MiAlgo & CASIA LCVG & HIT-IIL
    \tabularnewline
    \includegraphics[width=0.24\linewidth]{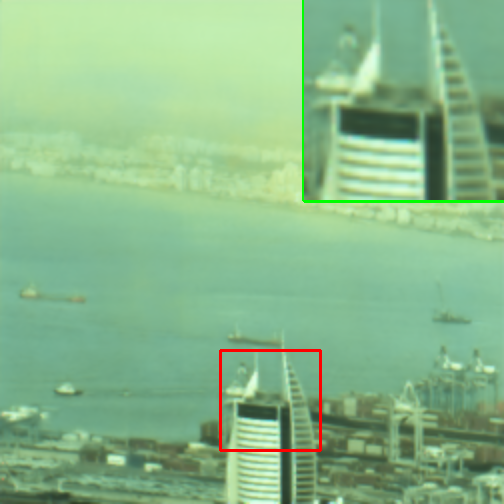} &
    \includegraphics[width=0.24\linewidth]{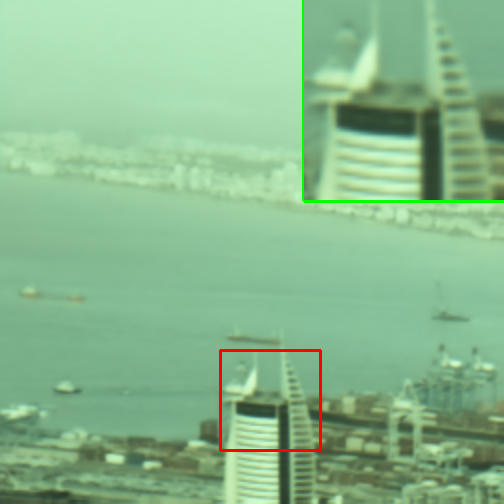} &
    \includegraphics[width=0.24\linewidth]{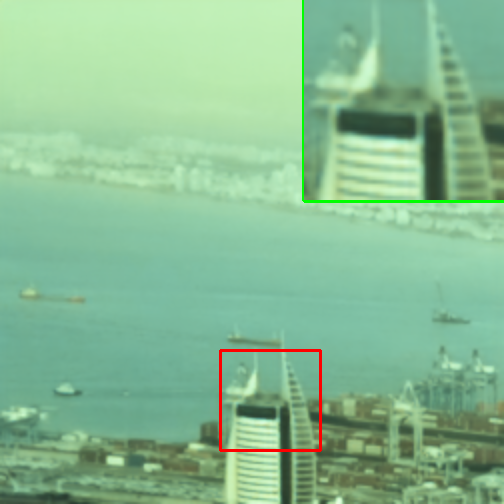} &
    \includegraphics[width=0.24\linewidth]{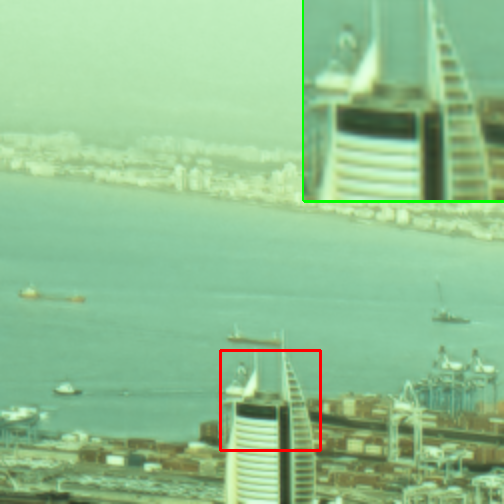}
    \tabularnewline
    SenseBrains & CSˆ2U & HiImage & 0noise
    \vspace{10mm}
    \tabularnewline
    \includegraphics[width=0.24\linewidth]{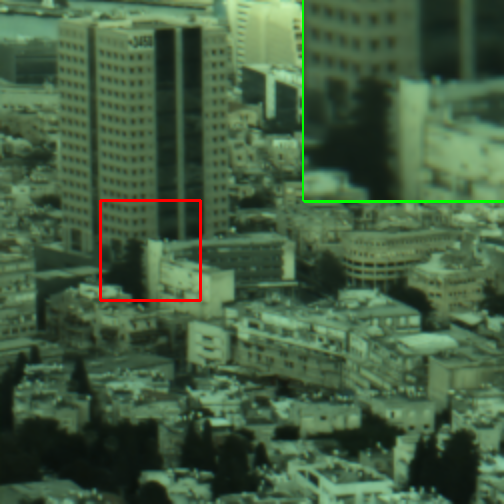} &
    \includegraphics[width=0.24\linewidth]{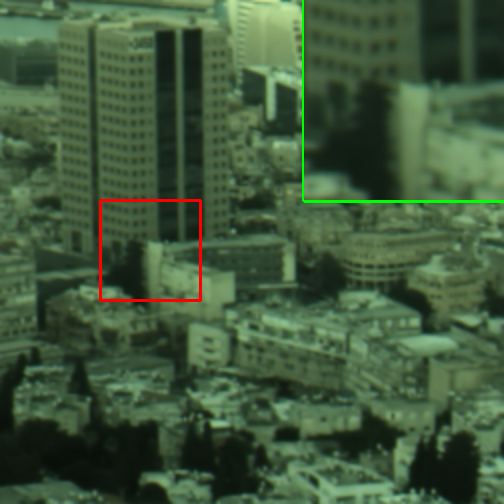} &
    \includegraphics[width=0.24\linewidth]{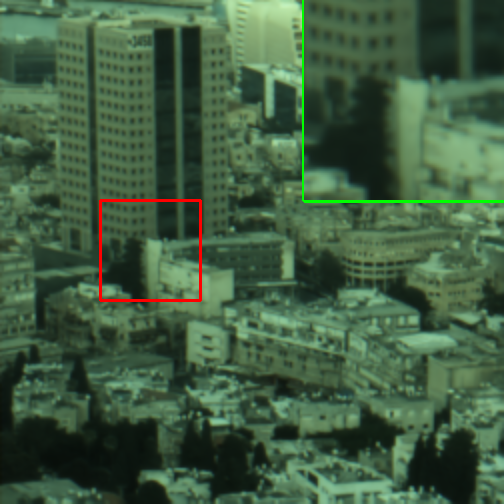} &
    \includegraphics[width=0.24\linewidth]{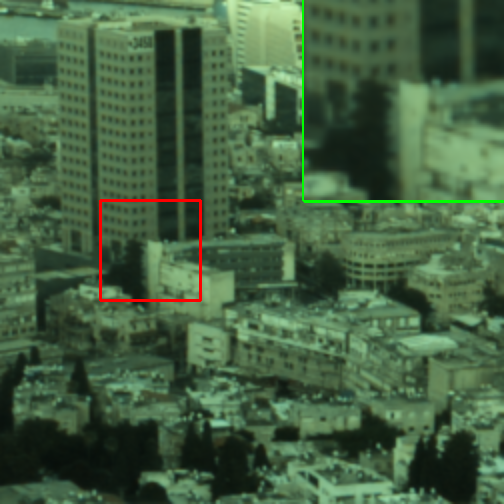}
    \tabularnewline
    NOAHTCV & MiAlgo & CASIA LCVG & HIT-IIL
    \tabularnewline
    \includegraphics[width=0.24\linewidth]{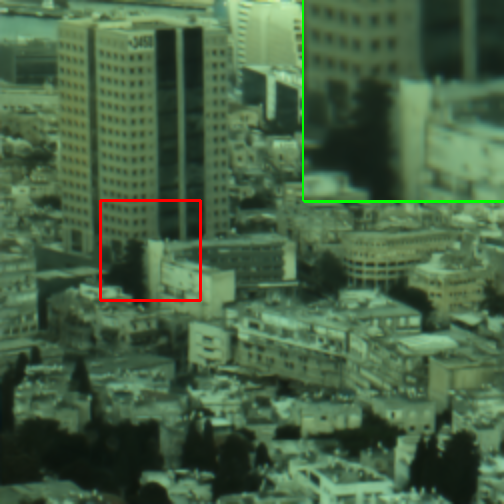} &
    \includegraphics[width=0.24\linewidth]{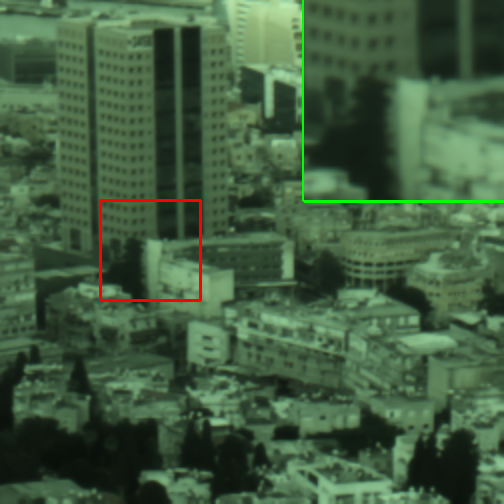} &
    \includegraphics[width=0.24\linewidth]{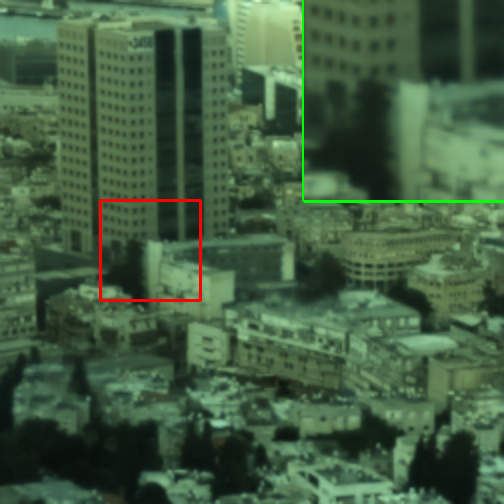} &
    \includegraphics[width=0.24\linewidth]{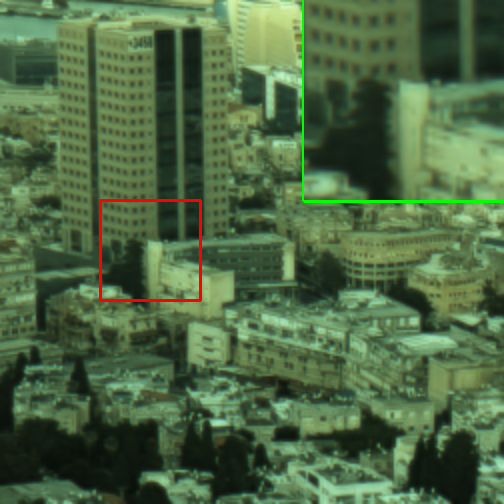}
    \tabularnewline
    SenseBrains & CSˆ2U & HiImage & 0noise
    \tabularnewline
    \end{tabular}
    \caption{Qualitative Comparison of the best methods on the Track 1 (S7). As we can see, these methods can recover detailed RAW images without artifacts.}
    \label{fig:quali-results2}
\end{figure}

\clearpage
\section{Appendix 2: Teams and affiliations}
\label{app:affiliations}

\bigskip
\noindent
\textbf{NOAHTCV}
\small
\bigskip


\noindent
\textbf{Members:} Yibin Huang, Jingyang Peng, Chang Chen, Cheng Li, Eduardo P{\'e}rez-Pellitero, and Fenglong Song

\smallskip
\noindent
\textbf{Affiliations:} Huawei Noah’s Ark Lab

\smallskip
\noindent
\textbf{Contact:} \small{songfenglong@huawei.com}\\

\bigskip
\noindent
\textbf{MiAlgo}
\small
\bigskip

\noindent
\textbf{Members:} Furui Bai, Shuai Liu, Chaoyu Feng, Xiaotao Wang, Lei Lei

\smallskip
\noindent
\textbf{Affiliations:} Xiaomi Inc., China

\smallskip
\noindent
\textbf{Contact:} \small{baifurui@xiaomi.com}\\

\bigskip
\noindent
\textbf{CASIA LCVG}
\small
\bigskip

\noindent
\textbf{Members:} Yu Zhu, Chenghua Li, Yingying Jiang, Yong A, Peisong Wang, Cong Leng, Jian Cheng

\smallskip
\noindent
\textbf{Affiliations:} Institute of Automation, Chinese Academy of Sciences; MAICRO; AiRiA

\smallskip
\noindent
\textbf{Contact:} \small{zhuyu.cv@gmail.com}\\

\bigskip
\noindent
\textbf{HIT-IIL}
\small
\bigskip

\noindent
\textbf{Members:} Xiaoyu Liu, Zhicun Yin, Zhilu Zhang, Junyi Li, Ming Liu, Wangmeng Zuo

\smallskip
\noindent
\textbf{Affiliations:} Harbin Institute of Technology, China

\smallskip
\noindent
\textbf{Contact:} \small{liuxiaoyu1104@gmail.com}\\

\vspace{2.5mm}
\noindent
\textbf{SenseBrains}
\small
\bigskip

\noindent
\textbf{Members:} Jun Jiang, Jinha Kim

\smallskip
\noindent
\textbf{Affiliations:} SenseBrain Technology 

\smallskip
\noindent
\textbf{Contact:} \small{jinhakim@sensebrain.site, jinhakim@mit.edu }\\

\vspace{2.5mm}
\noindent
\textbf{${\rm \textbf{CS}}^{2}{\rm \textbf{U}}$}
\small
\bigskip

\noindent
\textbf{Members:} Yue Zhang, Beiji Zou

\smallskip
\noindent
\textbf{Affiliations:} School of Computer Science and Engineering, Central South University; Hunan Engineering Research Center of Machine Vision and Intelligent Medicine.

\smallskip
\noindent
\textbf{Contact:} \small{yuezhang@csu.edu.cn}\\

\bigskip
\noindent
\textbf{HiImage}
\small
\bigskip

\noindent
\textbf{Members:} Zhikai Zong, Xiaoxiao Liu

\smallskip
\noindent
\textbf{Affiliations:} Qingdao Hi-image Technologies Co.,Ltd (Hisense Visual Technology Co.,Ltd.)

\smallskip
\noindent
\textbf{Contact:} \small{zzksdu@163.com}\\

\bigskip
\noindent
\textbf{0noise}
\small
\bigskip

\noindent
\textbf{Members:} Juan Mar{\'i}n Vega, Michael Sloth, Peter Schneider-Kamp, Richard Röttger

\smallskip
\noindent
\textbf{Affiliations:} University of Southern Denmark, Esoft Systems

\smallskip
\noindent
\textbf{Contact:} \small{marin@imada.sdu.dk}\\

\bigskip
\noindent
\textbf{OzU VGL}
\small
\bigskip

\noindent
\textbf{Members:} Furkan Kınlı, Barış Özcan, Furkan Kıraç

\smallskip
\noindent
\textbf{Affiliations:} Özyeğin University

\smallskip
\noindent
\textbf{Contact:} \small{furkan.kinli@ozyegin.edu.tr}\\

\bigskip
\noindent
\textbf{PixelJump}
\small
\bigskip

\noindent
\textbf{Members:} Li Leyi

\smallskip
\noindent
\textbf{Affiliations:} Zhejiang University

\smallskip
\noindent
\textbf{Contact:} \small{lileyi@zju.edu.cn}\\

\bigskip
\noindent
\textbf{CVIP}
\small
\bigskip

\noindent
\textbf{Members:} SM Nadim Uddin, Dipon Kumar Ghosh, Yong Ju Jung

\smallskip
\noindent
\textbf{Affiliations:} Computer Vision and Image Processing (CVIP) Lab, School of Computing, Gachon University.

\smallskip
\noindent
\textbf{Contact:} \small{smnadimuddin@gmail.com}\\


\end{document}